\PassOptionsToPackage{dvipsnames,svgnames}{xcolor} % bmcart loads xcolor

\documentclass{bmcart}

\makeatletter
\def\@listiii{\leftmargin\leftmarginiii
              \labelwidth\leftmarginiii
              \advance\labelwidth-\labelsep
              \topsep\z@
              \parsep\z@
              \partopsep\z@
              \itemsep\topsep}
\makeatother

\usepackage{lmodern}

\usepackage{graphicx}
\usepackage{etoolbox}
\usepackage{amssymb}
\usepackage{amsmath}
\usepackage{mathtools} % \prescript
\usepackage{bm}
\usepackage{braket}
\usepackage{xspace}

\usepackage{tabularx}
\usepackage[colorlinks=true, urlcolor=teal,linkcolor=teal, citecolor=teal]{hyperref}
\usepackage{adjustbox}
\usepackage{isotope} % for \isotope
\usepackage{siunitx} % for \qty
\usepackage{import}

\usepackage{array}
\newcolumntype{L}[1]{>{\raggedright\let\newline\\\arraybackslash\hspace{0pt}}m{#1}}
\newcolumntype{C}[1]{>{\centering\let\newline\\\arraybackslash\hspace{0pt}}m{#1}}
\newcolumntype{R}[1]{>{\raggedleft\let\newline\\\arraybackslash\hspace{0pt}}m{#1}}

\usepackage{stmaryrd} % only for [[n,k,d]] quantum code brackets

\newcommand{\qgate}[1]{\textsf{#1}}

\def\mathendash{\text{--}}
\def\abs#1{\left|#1\right|}

\def\Evec{\mathord{\bm E}}
\def\dvec{\mathord{\bm d}}
\def\nvec{\mathord{\bm n}}

\ifdefined\unit\else
  \ifdefined\NewCommandCopy
    \NewCommandCopy\unit\si
  \else
    \NewDocumentCommand\unit{O{}m}{\si[#1]{#2}}
  \fi
\fi

\startlocaldefs

\newif\ifdraft
\draftfalse

\ifdraft
\newcommand{\note}[1]{ {\textcolor{orange} { **: #1 }}}
\newcommand{\jknote}[1]{ {\textcolor{blue} { ***Johannes: #1 }}}
\newcommand{\grnote}[1]{ {\textcolor{cyan} { ***Georg: #1 }}}
\newcommand{\slnote}[1]{ {\textcolor{blue!75!white} { ***Sebastian: #1 }}}
\newcommand{\tenote}[1]{ {\textcolor{purple} { ***ThE: #1 }}}
\newcommand{\tsnote}[1]{ {\textcolor{Salmon} { ***TS: #1 }}}
\long\def\tslnote#1{\begingroup\color{Salmon}***TS: #1\endgroup}
\newcommand{\fdnote}[1]{ {\textcolor{green!50!black} { ***FD: #1 }}}
\newcommand{\kwnote}[1]{ {\textcolor{orange} { ***Karen: #1 }}}
\newcommand{\mynote}[1]{ {\textcolor{magenta} { ***MY: #1 }}}
\newcommand{\jwnote}[1]{ {\textcolor{purple} { ***JW: #1 }}}
\newcommand{\dbnote}[1]{ {\textcolor{Periwinkle} { ***DB: #1 }}}

\newcommand{\TODO}[1]{\textcolor{green}{TODO: #1}}
\newcommand{\TBD}{\textcolor{red}{TBD}}
\newcommand{\status}[1]{~[#1]}

\else

\newcommand{\note}[1]{}
\newcommand{\slnote}[1]{}
\newcommand{\tenote}[1]{}
\newcommand{\jknote}[1]{}
\newcommand{\grnote}[1]{}
\newcommand{\fdnote}[1]{}
\newcommand{\mynote}[1]{}
\newcommand{\kwnote}[1]{}
\newcommand{\jwnote}[1]{}
\newcommand{\tsnote}[1]{}
\newcommand{\tslnote}[1]{}
\newcommand{\dbnote}[1]{}

\newcommand{\TODO}[1]{}
\newcommand{\TBD}{}
\newcommand{\status}[1]{}

\fi

\endlocaldefs

\begin{document}
\begin{frontmatter}
\begin{fmbox}

  \dochead{Research}

  \title{Ion-Based Quantum Computing Hardware: Performance and End-User Perspective}

%%%%%%%%%%%%%%%%%%%%%%%%%%%%%%%%%%%%%%%%%%%%%%
%%                                          %%
%% Enter the authors here                   %%
%%                                          %%
%% Specify information, if available,       %%
%% in the form:                             %%
%%   <key>={<id1>,<id2>}                    %%
%%   <key>=                                 %%
%% Comment or delete the keys which are     %%
%% not used. Repeat \author command as much %%
%% as required.                             %%
%%                                          %%
%%%%%%%%%%%%%%%%%%%%%%%%%%%%%%%%%%%%%%%%%%%%%%

%% example from template \author[
%   addressref={aff1},                   % id's of addresses, e.g. {aff1,aff2}
%   corref={aff1},                       % id of corresponding address, if any
%   noteref={n1},                        % id's of article notes, if any
%   email={jane.e.doe@cambridge.co.uk}   % email address
%]{\inits{JE}\fnm{Jane E} \snm{Doe}}
%\author[
%   addressref={qutac},
%   email={info@qutac.de}
%]{\inits{QUTAC}\fnm{}\snm{Quantum Technology and Application Consortium – QUTAC}}
\author[
   addressref={bosch},
   email={Thomas.Strohm@de.bosch.com}
]{\inits{TS}\fnm{Thomas} \snm{Strohm}}
 \author[
    addressref={siemens},
    email={karen.wintersperger@siemens.com}
 ]{\inits{KW}\fnm{Karen} \snm{Wintersperger}}
\author[
   addressref={trumpf},
   email={florian.dommert@trumpf.com}
]{\inits{FD}\fnm{Florian} \snm{Dommert}}
\author[
   addressref={trumpf},
   email={daniel.basilewitsch@trumpf.com}
]{\inits{DB}\fnm{Daniel} \snm{Basilewitsch}}
\author[
   addressref={lhind},
   email={georg.reuber@lhind.dlh.de}
]{\inits{GR}\fnm{Georg} \snm{Reuber}}
\author[
    addressref={sap},
    email={andrey.hoursanov@sap.com}
 ]{\inits{AH}\fnm{Andrey} \snm{Hoursanov}}
\author[
   addressref={merck},
   email={thomas.ehmer@merckgroup.com}
]{\inits{TE}\fnm{Thomas} \snm{Ehmer}}
\author[
   addressref={basf},
   email={davide.vodolat@basf.com}
]{\inits{DV}\fnm{Davide} \snm{Vodola}}
\author[
    addressref={infineon},
    email={sebastian.luber@infineon.com}
]{\inits{SL}\fnm{Sebastian} \snm{Luber}}
% \author[
%    addressref={bmw},
%    email={Johannes.Klepsch@bmw.de}
% ]{\inits{JK}\fnm{Johannes} \snm{Klepsch}}
% \author[
%    addressref={siemens},
%    email={wolfgang.mauerer@siemens.com}
% ]{\inits{WM}\fnm{Wolfgang} \snm{Mauerer}}

% \author[
%    addressref={telekom},
%    email={Ming.Yin@telekom.de}
% ]{\inits{MY}\fnm{Ming} \snm{Yin}}

%%%%%%%%%%%%%%%%%%%%%%%%%%%%%%%%%%%%%%%%%%%%%%
%%                                          %%
%% Enter the authors' addresses here        %%
%%                                          %%
%% Repeat \address commands as much as      %%
%% required.                                %%
%%                                          %%
%%%%%%%%%%%%%%%%%%%%%%%%%%%%%%%%%%%%%%%%%%%%%%

%\address[id=qutac]{%                           % unique id
%  \orgname{QUTAC}, % university, etc
%  %\street{},                     %
%  %\postcode{}                                % post or zip code
%  %\city{London},                              % city
%  \cny{Germany}                                    % country
%}
\address[id=bosch]{%
  \orgname{Robert Bosch GmbH, Corporate Research},
  \street{Robert-Bosch-Campus 1},
  \postcode{71272}
  \city{Renningen},
  \cny{Germany}
}
\address[id=siemens]{%
  \orgname{Siemens AG},
  \street{Otto-Hahn-Ring 6},
  \postcode{81739}
  \city{München},
  \cny{Germany}
}
\address[id=trumpf]{%
  \orgname{TRUMPF SE + Co. KG},
  \street{Johann-Maus-Straße 2},
  \postcode{71254}
  \city{Ditzingen},
  \cny{Germany}
}
\address[id=lhind]{%
  \orgname{Lufthansa Industry Solutions GmbH \& Co. KG},
  \street{Sch\"utzenwall 1},
  \postcode{22844}
  \city{Norderstedt},
  \cny{Germany}
}
\address[id=sap]{%
  \orgname{SAP SE},
  \street{Dietmar-Hopp-Allee 16},
  \postcode{69190}
  \city{Walldorf},
  \cny{Germany}
}
\address[id=merck]{%
  \orgname{Merck KGaA},
  \street{Frankfurterstr. 250},
  \postcode{64293}
  \city{Darmstadt},
  \cny{Germany}
}
%\address[id=bmw]{%
%  \orgname{BMW AG},
%  \street{Bremer Str. 6},
%  \postcode{80807}
%  \city{Munich},
%  \cny{Germany}
%}
%\address[id=telekom]{%
%  \orgname{Deutsche Telekom},
%  \street{Winterfeldtstraße 21},
%  \postcode{ 10781},  
%  \city{Berlin},
%  \cny{Germany}
%}
\address[id=basf]{%
  \orgname{BASF Digital Solutions},
  \street{Pfalzgrafenstr. 1},
  \postcode{67061}
  \city{Ludwigshafen am Rhein},
  \cny{Germany}
}
\address[id=infineon]{%
  \orgname{Infineon Technologies AG},
  \street{Am Campeon 1-15},
  \postcode{85579}
  \city{Neubiberg},
  \cny{Germany}
}

%%%%%%%%%%%%%%%%%%%%%%%%%%%%%%%%%%%%%%%%%%%%%%
%%                                          %%
%% Enter short notes here                   %%
%%                                          %%
%% Short notes will be after addresses      %%
%% on first page.                           %%
%%                                          %%
%%%%%%%%%%%%%%%%%%%%%%%%%%%%%%%%%%%%%%%%%%%%%%

\begin{artnotes}
%\note{Sample of title note}     % note to the article
%\note[id=n1]{Equal contributor} % note, connected to author
\end{artnotes}

\end{fmbox}% comment this for two column layout

\begin{abstractbox}

\begin{abstract}
  %%%%%%%%%%%%%%%%%%%%%%%%%%%%%% new abstract (start)
  %\dbnote{Suggestion for a new abstract:}\tsnote{Looks good. Fine for me. I changed ``overview over'' to ``overview of'', the latter seems to be preferred.}\\
  This is the second paper in a series of papers providing an overview of  different quantum computing hardware platforms from an industrial end-user perspective.
  It follows our first paper on neutral-atom quantum computing~\cite{Wintersperger2023}.

  In the present paper, we provide a survey on the current state-of-the-art in trapped-ion quantum computing, taking up again the perspective of an industrial end-user.
  To this end, our paper covers, on the one hand, a comprehensive introduction to the physical foundations and mechanisms that play an important role in operating a trapped-ion quantum computer.
  On the other hand, we provide an overview of the key performance metrics that best describe and characterise such a device's current computing capability.
  These metrics encompass performance indicators such as qubit numbers, gate times and errors, native gate sets, qubit stability and scalability as well as considerations regarding the general qubit types and trap architectures.
  In order to ensure that these metrics reflect the current state of trapped-ion quantum computing as accurate as possible, they have been obtained by both an extensive review of recent literature and, more importantly, from discussions with various quantum hardware vendors in the field.
  We combine these factors and provide~-- again from an industrial end-user perspective~-- an overview of what is currently possible with trapped-ion quantum computers, which algorithms and problems are especially suitable for this platform, what are the relevant end-to-end wall clock times for calculations, and what might be possible with future fault-tolerant trapped-ion quantum computers.
\end{abstract}


\end{abstractbox}

\end{frontmatter}

%%%%%%%%%%%%%%%%%%%%%%%%%%%%%%%%%%%%%%%%%%%%%%%%%%%%%%%%%%%%%%%%%%%%%%%%%%%%

\tableofcontents

\section{Introduction\status{Daniel,OK}}
\label{sec:introduction}

Quantum computing hardware has improved tremendously in recent years.
It already showcases today~-- despite its still early stage~-- what might be feasible with such machines once they unfold their full potential.
Quantum computing promises applications in various fields, covering classical cryptography~\cite{SIAM.26.1484}, optimisation~\cite{Abbas2023}, quantum chemistry and material science~\cite{Bauer2020}, and finance~\cite{Herman2023} to name a few.
However, despite considerable progress in hardware development, applications for existing quantum computers have so far been mostly limited to proof-of-principle examples~\cite{Arute2019, PRL.127.180501, Zhong2020, PRL.127.180502, Bluvstein2023}.
Albeit some of these examples have demonstrated advantage over classical computers~-- in terms of required computational resources~-- the tasks that have been solved in these examples are, unfortunately, of little interest from an industry point of view.
This is primarily owed to the fact that these examples needed to be chosen and tailored to be executable given the limited capabilities of current quantum hardware.
While these limitations have diverse origins, the presence of various sources of errors~-- be it due to inevitable environmental interaction or noisy operational hardware~-- and the technical challenge to scale up qubit numbers are two of the main limitations.
Achieving sufficiently small error rates and sufficiently large qubit numbers will likely enable quantum error correction and allow for fault-tolerant quantum computing in the future, with first steps into this direction being completed recently~\cite{Krinner2022, Sivak2023, Bluvstein2023, da_Silva_Ryan-Anderson_Bello-Rivas_Chernoguzov_Dreiling_Foltz_Gaebler_Gatterman_Hayes_Hewitt_etal.}.
However, current quantum hardware is not there yet.
In consequence, this ultimately prohibits more complex application, like tackling real-world problems of economic and industrial interest, for now.
Nevertheless, even in the current era of noisy intermediate-scale quantum (NISQ) devices~\cite{Preskill_2018}, there has already been significant interest from academia and industry to explore the potential of quantum computing.

There are various different physical platforms to build a quantum computer, each of them having their individual strengths and challenges.
Some of the most promising approaches include neutral atoms~\cite{Wintersperger2023}, trapped ions~\cite{Bruzewicz2019}, photons~\cite{OBrien2007}, superconducting circuits~\cite{Blais+2021}, spins in semiconductors~\cite{Kloeffel2013} and colour centers in diamond~\cite{Pezzagna2021}.
While it is not yet clear when or whether any platform will reach the point of being able to run useful, fault-tolerant quantum computations, it is clear that up to that point~-- and maybe even beyond~-- some applications will naturally benefit from platform-specific features which renders them a perfect fit for that particular platform.
In order to exploit such features in the current NISQ era and beyond, it is thus inevitable for an end-user to have a basic overview and understanding of each platform's working principle and key performance metrics.
For a very approachable, platform-independent overview, also to non-experts, we recommend e.g.~\cite{ezratty_2023}.

The paper at hand targets an end-user audience from industry and provides an overview of the most relevant properties and current state-of-the-art for one particular quantum computing platform, namely trapped ions. 
It is the second paper in a series which intends to provide such an overview for different quantum computing platforms. For an introduction to neutral-atom quantum computing, we refer the interested reader to this series' first paper~\cite{Wintersperger2023}.

The remainder of the paper is organised as follows.
The physical foundations underlying trapped-ion quantum computers are explained in Sec.~\ref{sec:foundations}.
This encompasses the trapping of ions, definition of qubits, their control and readout as well as interaction between qubits.
Subsequently, Sec.~\ref{sec:platform_overview.tex} provides an overview of the different architectural approaches to build a trapped-ion quantum computer from the physical principles introduced before.
This overview is embedded in the general discussion on how to scale up qubit numbers and correct for errors in order to ultimately enable fault-tolerant quantum computing in the future.
Section~\ref{subsec:explanation_criteria} summarises the state-of-the-art of present-day trapped-ion quantum hardware in terms of performance metrics.
To which extent trapped-ion quantum computers have already been used nowadays for academic or industrial calculations is discussed in Sec.~\ref{subsec:application_overview.tex}.
The last Sec.~\ref{sec:discussion} summarises the main advantages and disadvantages of trapped-ion quantum computers, also in comparison with other prominent quantum computing platforms, and discusses examples of applications that profit particularly from the characteristics of this hardware platform.

\section{Physical foundations of the ion trap platform\status{Intro text: KW, OK}}
\label{sec:foundations}

To better understand the properties and performance of trapped-ion quantum computers, we start with a brief overview of the fundamental physics and resulting implementations. The foundations have been laid by Cirac and Zoller in 1995 in their seminal paper~\cite{cirac_zoller_1995_PhysRevLett.74.4091}.  Since then, the field has made a lot of progress, which is documented in several technical reviews, e.g.,~\cite{Leibfried2003,blatt-wineland-2008,haffner_2008,Wineland_2009,Schindler_2013,Bruzewicz2019}. Many of the related technologies existed before and/or are used also in different contexts: ion trapping, laser cooling, readout via state-dependent fluorescence in high-precision spectroscopy and atomic clocks.

\subsection{Overall structure and operation\status{KW,OK}}
\label{sec:quantum_computing_sequence}

The architecture of most trapped-ion quantum computers consists of a vacuum chamber containing the trap, which is a small device or a micro-fabricated chip.  The trap contains electrodes that create a static electric field, which is combined either with an additional oscillating electric or a static magnetic field, depending on the trap type being used. In this field, the ions are trapped in linear chains or two-dimensional arrangements. A trapped-ion quantum computer can operate at room temperature. Depending on the specific trap and control setup being used, temperatures of ion traps can be as high as $\approx \qty{100}{\degreeCelsius}$~\cite{Pogorelov_2021}, arising, e.g., from the power of the driving signals. To reduce thermal noise, ion traps are nevertheless often cooled to cryogenic temperatures~\cite{Pino_Dreiling_Figgatt_Gaebler_Moses_Allman_Baldwin_Foss-Feig_Hayes_Mayer_et_al._2021} using refrigerators, which in addition improves the vacuum. 

Overall, trapped-ion quantum computing setups consist of various components, also including lasers and control devices. Nevertheless, they have already been miniaturised to rack size~\cite{Pogorelov_2021}, fitting the complete quantum computer into two racks. Quantum computers will most likely be used in a hybrid setup alongside classical computers. Thus, the infrastructure requirements for a particular quantum computing platform need to be considered in terms of integrating it, for instance, within a classical supercomputer or installing it at the user's facility such as a production shop floor (in the usual case, however, quantum computers will likely be used remotely).

To prepare and execute a quantum computation, a sequence of operations has to be performed, which is described in the following.  First, the trap has to be loaded with ions. To this end, one starts with the generation of atomic vapour from a thermal source of bulk material, although other techniques such as ablation can be employed~\cite{Pogorelov_2021}. In most cases, the setup consists of a single vacuum chamber that contains the atomic source and the trap. However, also other approaches~\cite{quantinuum2023_Moses_Baldwin_Allman_Ancona_Ascarrunz_Barnes_Bartolotta_Bjork_Blanchard_Bohn_etal._2023} exist where a separate vacuum chamber is used to create and precool the atomic vapour before it is transferred into the primary vacuum chamber with the trap. The latter procedure allows to maintain a lower pressure in the trapping cell.

Initially, neutral atoms from the atomic vapour are then ionised by absorption of laser light, often with two laser beams~\cite{Olmschenk_2007, Pogorelov_2021} to provide the necessary total ionisation energy. The ionisation takes place close to the trap centre such that the generated ions are confined by the trapping potential afterwards. To prepare them for computations, the motion of the trapped ions is laser cooled, using Doppler cooling and subsequent Raman sideband cooling~\cite{wright_benchmarking_2019, quantinuum2023_Moses_Baldwin_Allman_Ancona_Ascarrunz_Barnes_Bartolotta_Bjork_Blanchard_Bohn_etal._2023}. In some cases also polarisation gradient cooling~\cite{Ejtemaee_2017} is employed prior to the sideband cooling~\cite{Pogorelov_2021}.

After trapping and cooling, the ions are initialised to the $\ket{0}$ state by optical pumping, as the starting point for computation. Quantum gates are executed by application of laser beams or microwaves. To realise single qubit gates, laser beams are typically directed towards single ions. However, as an alternative~\cite{Lekitsch2017}, it is also possible to use a global microwave pulse acting on all ions simultaneously but using a tailored frequency such that it effectively addresses only a single ion. In the latter scheme, the qubit frequencies of the ions vary due to application of a magnetic gradient. For a two-qubit gate, the vibrational modes (motion) of the ion chain are utilised such that the qubit levels are coupled to the same motional mode and thus effectively coupled to each other.  This is either achieved by using laser light or, as described before, by combining microwave pulses with a magnetic gradient.

\begin{figure}
    \centering
    \includegraphics[scale=0.8]{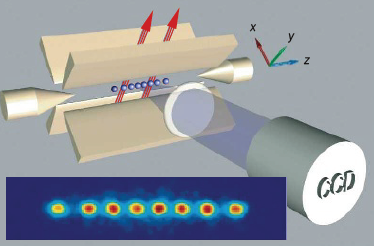}%
    \caption{Schematics of a trap with some trapped ions, laser beams to implement gates or cause fluorescence for readout and a CCD camera to detect the fluorescence.  Everything is inside of a ultra high vacuum chamber.  The inset shows a photography of a fluorescing ion chain.  Source:~\cite{blatt-wineland-2008}.}
    \label{fig:trap-scheme}
\end{figure}

The final state of the ions is read out by exciting fluorescence and collecting the light onto a camera. Depending on its state, each ion appears as a bright or dark spot. Some architectures allow reading the state of a qubit at arbitrary time without disturbing other qubits, followed by re-initialisation of the qubit (mid-circuit measurement)~\cite{Pino_Dreiling_Figgatt_Gaebler_Moses_Allman_Baldwin_Foss-Feig_Hayes_Mayer_et_al._2021}. The individual steps of the computing sequence are described in more detail in the following sections. Figure~\ref{fig:trap-scheme} illustrates the entire ion-trap setup with trapping electrodes, ion/qubit control fields and readout mechanisms schematically.

\subsection{Implementing qubits\status{Karen,OK}}
\label{sec:implementing}

The choice of using ions in quantum computing is driven by several factors~\cite{Bruzewicz2019}. Firstly, ions are charged and can be trapped by electromagnetic fields, which enables their confinement in a small volume. This confinement facilitates the precise control and manipulation of individual qubits, which is essential for implementing quantum gates. Additionally, ions can be coupled over long distances, up to several tens of microns, which is important for implementing multi-qubit gate operations. Coupling ions via their common motional mode allows for an effective all-to-all connectivity avoiding the insertion of \qgate{SWAP} gates and facilitating error correction. Trapped ions are well isolated from the environment, which allows for high fidelities of gate operations and measurement. Ion qubits also exhibit long coherence times leading to ratios of coherence times to gate times of more than~$\approx 10^3$ (see Tab.~\ref{tab:Parameters_1}).  This is a rough indicator of the depth of quantum circuits that can be carried out with tolerable error. Compared to macroscopic qubits such as superconducting circuits, ions are fundamentally all identical by nature, although spatially-varying external perturbations can lead to effective differences in coherence time and qubit frequencies. 
%Another advantage of using trapped ions is the ability to hybridise several different ion types, such as calcium and strontium, to obtain their respective benefits, such as fast gate operations for calcium and stability for strontium.

Furthermore, the simple electron configuration of the chosen ion elements (usually closed shells plus an additional electron in the $s$-state) results in short-lived excitation levels from the ground state, which allows for laser cooling. This leads to stable basic energy states corresponding to $\ket{0}$ and excited energy states corresponding to $\ket{1}$, which facilitates their use as qubits for implementing quantum gates and operations.\footnote{In this paper, we use the notations~$\ket0, \ket1$ for the computational basis states of a qubit as well as~$\ket g, \ket e$ for the ground and excited state, respectively. We identify~$\ket0\equiv\ket g$ and~$\ket1\equiv\ket e$.}

In this section we first describe the different types of ions than can be~$\ket{0}$ and~$\ket{1}$, which are called the \textit{computational basis states}.
Subsequently, the trapping of ions as well as other steps from the computation sequence related to the physics of the qubits such as cooling, initialisation and readout are explained in more detail. 

\subsubsection{Qubit flavours\status{KW,TS,OK}}
\label{sec:qubits}

\paragraph{Energy level diagram}

The ions in a trapped-ion quantum computer are usually singly ionised group-II or group-II-like atoms. The electron configuration of these ions is~$s^1$, they have completely filled shells plus one electron in a rotation symmetric $s$-state.  In comparison to atoms with a different electron configuration, the term scheme of these ions is relatively simple and similar to the one of the hydrogen atom.

\begin{figure}
    \centering
    \includegraphics[width=0.95\textwidth]{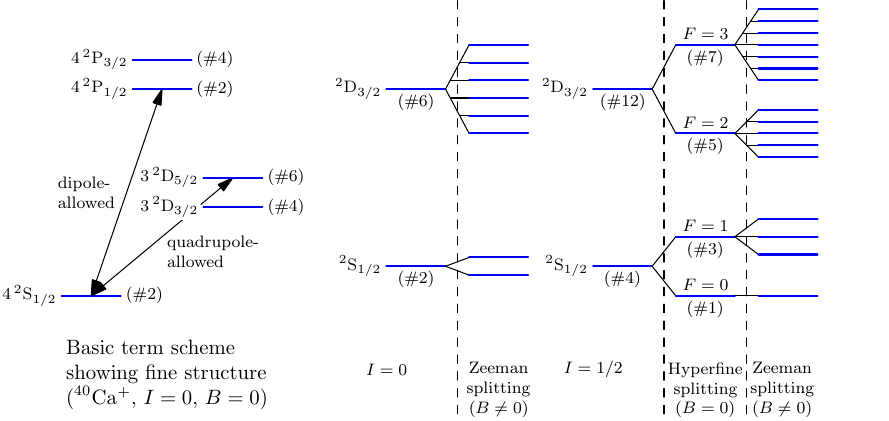}
    %{img/term-scheme-group-II.png}
    \caption{Typical energy level diagram for singly ionised group-II atoms.  Left side: Energy levels of~$\isotope[40]{Ca}^+$, which has zero nuclear spin, at~$B=0$. The lines are the \textit{fine structure}.  The label~$\#n$ at the lines gives their multiplicity.  Right side: Fine structure for an~$I=0$ nucleus vs.\ an~$I=1/2$ nucleus with \textit{hyperfine} structure (due to the nuclear spin) at~$B=0$ and at~$B\neq0$.  With~$B\neq0$, all degeneracy is lifted.}
    \label{fig:term-scheme-group-II-ions}
\end{figure}

As an example, Fig.~\ref{fig:term-scheme-group-II-ions}, left side, shows the energy level diagram of~$\isotope[40]{Ca}^+$, which in the ground state~$4S_{1/2}$ has an electron configuration of~$1s^2\,2s^2 p^6\,3s^2 p^6\,4s^1$ with an unpaired electron in the~$4s$ shell. The $3d$ subshell is not occupied. This energy level diagram is typical for an atom with one unpaired~$s$ electron and vanishing nuclear spin~$I=0$.

%In the~$4P_J$ states, the unpaired electron occupies the~$4p$ subshell.  In the lower-lying~$3D_J$ states, the electron occupies the~$3d$ subshell (not in shell~4 but in shell~3, which is the reason why the~$4P_J$ states have higher energy than the~$3D_J$ states).  This term scheme is typical for an atom with one unpaired~$s$ electron and no nuclear spin~$I$.

It is notable that an electron in the~$4p$ shell has a higher energy than an electron in the~$3d$ shell, which again has a higher energy than an electron in the~$4s$ shell. This is the reason why the~$3D$ energy levels lie in between the~$4S$ and the~$4P$ energy levels.

For a non-zero nuclear spin, $I\neq0$, the energy level diagram shows the hyperfine structure: the levels split into two different subterms. In Fig.~\ref{fig:term-scheme-group-II-ions}, right side, the example of a nucleus with spin~$I=1/2$ is shown. With a vanishing magnetic field~$B=0$, there is still degeneracy (indicated by the~$\#n$ notation in the figure), but this is lifted in the presence of a magnetic field~$B \neq 0$.

\paragraph{Qubit variants and trade-offs in selection}

Defining a qubit corresponds to selecting two states out of the ion's spectrum.  There are basically four different types of qubits~\cite{Bruzewicz2019}.
%
% Zeeman: energy levels would be degenerate for B=0
% Hyperfine: energy levels would be degenerate for I=0
% Fine structure: same orbital angular momentum (e.g., D), but different J
% Optical: different orbital angular momentum
\begin{itemize}
    \item \textit{Zeeman qubits}: Levels of the same degenerate multiplet are split using a bias magnetic field~$B \neq 0$. This yields transition frequencies in the order of~\qtyrange{1}{10}{MHz}. The ground state of atoms with zero nuclear spin is typically used.
    \item \textit{Hyperfine qubits}: Levels of the same fine structure multiplet, that are split by a non-zero nuclear spin (hyperfine structure) are used (e.\,g. the~$F=0,1$ multiplets of~${}^2S_{1/2}$ for a nucleus with~$I=1/2$ and $F$ the total angular momentum). If the levels resulting from hyperfine splitting are still degenerate, a non-zero bias magnetic is used to remove the remaining degeneracy.  Typical transition frequencies are in the order of~\qtyrange{1}{10}{GHz}.
    \item \textit{Optical qubits}: States with a different orbital angular momentum~$L$ are used (e.\,g., ${}^2S_{1/2}$ and~${}^2D_{5/2}$ in a singly ionised group-II atom).  The transition frequencies are in the order of~\qtyrange{100}{1000}{THz}.
    \item \textit{Fine-structure qubits} are also possible, but, to the best of our knowledge, they are not used in practice.  In this case, the states would have the same orbital angular momentum~$L$ but different total angular momentum~$J$ (e.\,g., ${}^2D_{3/2}$ and~${}^2D_{5/2}$ in a singly ionised group-II atom). The transition frequencies are in the order of~\qty{1}{THz}.
\end{itemize}

The selection of a particular qubit requires several considerations and involves trade-offs.  For instance, the lifetime of the qubit should be high and therefore, the decay time of the upper qubit level should be low.  Lifetimes are limited by spontaneous emission, which is proportional to the cube of the transition frequency and the transition matrix element.  Therefore, one either needs small transition frequencies (Zeeman or hyperfine qubits) or higher order transitions (i.e., relying on e.g.\ exclusively quadrupole- or octupole-allowed transitions between qubit levels). However, inducing transitions in these long-lifetime qubits is difficult and large radiation intensities are needed. Transitions for optical qubits are induced with (optical) laser pulses.  If the frequencies are not too high, this is very convenient. For Zeeman or hyperfine qubits, one needs radio frequency (RF) or microwave pulses.

It is very difficult to address one particular ion in a chain with this long-wavelength radiation, as it cannot be focused tightly enough. Therefore, the transitions are often realised via the difference of two laser frequencies (see below).  Zeeman qubits in general are advantageous, but the energy of Zeeman levels changes with the magnetic field and thus they are very sensitive to magnetic field fluctuations.

It is important to note that none of the two qubit levels must be degenerate. Therefore, a bias magnetic field is usually used to lift the degeneracy of the multiplets.

\paragraph{Zeeman qubits}
% Oxford ionics (Srinivas_etal._2023), NeQxt (Hilder2022,Ruster2016,Poschinger_2009)

This type of qubit uses two sublevels of the atomic ground state.  For atomic species with nuclear spin $I=0$, they are part of the same fine structure manifold. As there is no nuclear spin, the energy levels are given by the coupling of orbital angular momentum~$L$ and electron spin~$S$, whereas $L=0$ for the ground state. The energy degeneracy between the levels with different electron spin magnetic quantum numbers~$m_S=\pm1/2$ is lifted by applying an external magnetic field $B \neq 0$, leading to typical energy differences in the MHz range. The relatively simple level structure allows for straightforward implementation of state initialisation, optical pumping, and cooling.

A prominent example of a Zeeman qubit is $\isotope[40]{Ca}^+$ (see Fig.~\ref{fig:term-scheme-group-II-ions}), where the levels~$\ket{4{}^2S_{1/2}, m_S=\pm1/2}$
%$\ket{4{}^2S_{1/2}, m_J = +1/2}$ and $\ket{4^2S_{1/2}, m_J = -1/2}$ 
are employed as qubit states~\cite{Srinivas_etal._2023, Hilder2022,Ruster2016,Poschinger_2009}. Since the energy difference is very small, the levels are coupled by stimulated Raman transitions with two lasers at $379\,\unit{nm}$ wavelength being slightly detuned from the $S_{1/2}$-$P_{1/2}$ transition such that their frequency difference matches the qubit frequency. To read out the state of the qubit, population from one qubit level is transferred to the metastable~$D_{5/2}$ state and state-dependent fluorescence of the $S_{1/2}$-$D_{5/2}$ quadrupole transition at $729\,\unit{nm}$ is detected. 

A disadvantage of Zeeman qubits is that fluctuations of the magnetic field have a detrimental effect to them and must be precisely controlled.

\paragraph{Hyperfine qubits}
%IonQ (egan_2021), Quantinuum (Olmschenk_Younge_Moehring_Matsukevich_Maunz_Monroe_2007), EleQtron(Bassler2023)
 
In a nucleus with non-zero nuclear spin $I \neq 0$, the total angular momentum  of the electrons couples with the nuclear spin and, for the ions with one unpaired electron, two hyperfine multiplets with $F=0$ and $F=1$ are formed. The remaining degeneracy of the~$F=1$ multiplet is usually lifted via the Zeeman effect by applying a small bias magnetic field~$B \neq 0$ that defines the quantization axis. The qubit levels are commonly chosen as states within the hyperfine multiplets with $F=0$ and $F=1$. The transition frequencies are by some orders of magnitude higher than for a Zeeman qubit and lie in the GHz range. Hyperfine qubits exhibit long lifetimes and coherence times and are less sensitive to magnetic field noise than Zeeman qubits, but in general have a more complicated level structure. 

An example is the $\isotope[171]{Yb}^+$ ion, which has~$I=1/2$ and whose ground state multiplet~${}^2S_{1/2}$ has a multiplicity of four and splits into two hyperfine terms~$F=0$ and~$F=1$~\cite{Olmschenk_Younge_Moehring_Matsukevich_Maunz_Monroe_2007}.  A non-zero bias magnetic field lifts the degeneracy of the~$F=1$ triplet (see Fig.~\ref{fig:term-scheme-group-II-ions}, right side). The states with~$m_F=0$ ($m_F$ denotes the magnetic quantum number associated with $F$), also called clock states, do not depend on the magnetic field in first order and are often used as qubit states~\cite{Olmschenk_Younge_Moehring_Matsukevich_Maunz_Monroe_2007, egan_2021}. As the transition frequency also lies in the microwave range, direct addressing of qubits by focusing a single radiation beam is not possible. Thus, the qubit levels are coupled by Raman transitions to the $^2P_{1/2}$ level using two laser beams in the same manner as described above for Zeeman qubits. Another approach is to directly couple the qubit levels in $\isotope[171]{Yb}^+$ using microwave pulses~\cite{Bassler2023}. To allow for single qubit addressing, the upper qubit level is chosen as one of~$\ket{F=1, m_F = 0,\pm1}$, and a spatial magnetic gradient along the ion chain is applied, leading to different transition frequencies for each qubit. Thus, a global microwave pulse with the corresponding frequency only interacts with the desired qubit. 

Although $\isotope[171]{Yb}^+$ ions are commonly used as qubits, they have the drawback that the laser frequencies needed for excitation lie in the UV range. An alternative which might be pursued in future devices is $\isotope[137]{Ba}^+$~\cite{IonQ_BA_announcement}, for which lasers in the visible range can be used which are easier to build and which also enables the use of photonic technologies. This can help to further increase the gate fidelities and obtain more reliable quantum computers. Also, $\isotope[137]{Ba}^+$ ions can be employed to create entangled states with photons at telecommunications wavelengths, enabling the building of quantum networks~\cite{saha_low-noise_2023}. 

\paragraph{Optical qubits}
%AQT (Pogorelov_2021), Oxford ionics (Srinivas_etal._2023) 

For optical qubits, the ground state~${}^2S_{1/2}$ and an excited state at an excitation energy corresponding to an optical frequency is typically used. The large transition frequency causes a large spontaneous emission rate and therefore a short lifetime.  This can be combated with a small transition rate given by the quadrupole transition to a~$D$ state (which, in the considered scheme, is available for~$\isotope{Ca}$ and heavier singly ionised group-II elements) or even by the octopole transition to an~$F$ state (which is available for~$\isotope{Ba}$ and heavier elements).  The smaller the transition rate, the slower is the speed of a gate (i.\,e., the Rabi frequency) or the higher is the optical power needed.

A typical ion used as an optical qubit is $\isotope[40]{Ca}^+$ (see Fig.~\ref{fig:term-scheme-group-II-ions}) with the states $\ket{3D_{5/2},m_J=-1/2}$ and~$\ket{4S_{1/2},m_J=-1/2}$ defining the qubit levels~\cite{Pogorelov_2021, Srinivas_etal._2023}. The transition is realised via electric quadrupole coupling with a wavelength of about $729\,\unit{nm}$.  The quality of laser light, light detection, and the optical elements is usually better in the IR and visible as in the UV.  This is another reason, why the transition frequencies should not be too large.

\subsubsection{Trapping, ion chains and vibrational modes\status{DV,TS,OK}}
\label{sec:trapping}

\def\orf{{\omega_\text{rf}}}  % use this for trap frequency

%\tenote{We can mention also the article \cite{Png_Hsu_Liu_Lin_Chang_2022}, the question is in how far we want to go deep for traps in general.}

\paragraph{Trap types}

To use ions as computational objects, precise control over them within a well-defined space is needed, which can be achieved by employing so-called ion traps. The basic idea of the ion trap goes back to the 1950s. It actually evolved from a mass-spectroscopic measurement device in high-energy physics~\cite{Paul1990}: the quadrupole mass filter. The inventors of the mass filter realised that they can also build a device that confines ions in a certain volume of space by applying a suitable spatial and temporal configuration of an electromagnetic field. A charged particle cannot be trapped solely with an electrostatic potential, as given by~\textit{Earnshaw's theorem}.\footnote{The reason can easily be seen: to trap a charged particle at the origin of the coordinate system, the potential in the vicinity of the origin would need to have the form~$V(x,y,z) = ax^2 + by^2 + cz^2$ with~$a,b,c>0$.  However, Laplace's equation for an electric field in free space requires~$a+b+c=0$.}

Two main approaches have been developed to store charged particles as ions: the \textit{Penning trap}, which is based on adding static magnetic fields, and the \textit{Paul trap}, which operates with a time-dependent electric field oscillating in the RF range. The latter one is mainly used within trapped-ion quantum computers and is also referred to as RF trap.\footnote{The complication with the Penning trap is that the ion chain would rotate and this complicates the handling~\cite{Bruzewicz2019}.} In the following we will sketch the physics of ion trapping using the Paul trap. We recommend the reviews~\cite{Paul1990, Wineland_Monroe_Itano_Leibfried_King_Meekhof_1997, Leibfried2003, Cho2015, Bruzewicz2019, Png_Hsu_Liu_Lin_Chang_2022}, which address this topic in more detail.

\begin{figure}
\centering
\includegraphics[width=.4\linewidth]{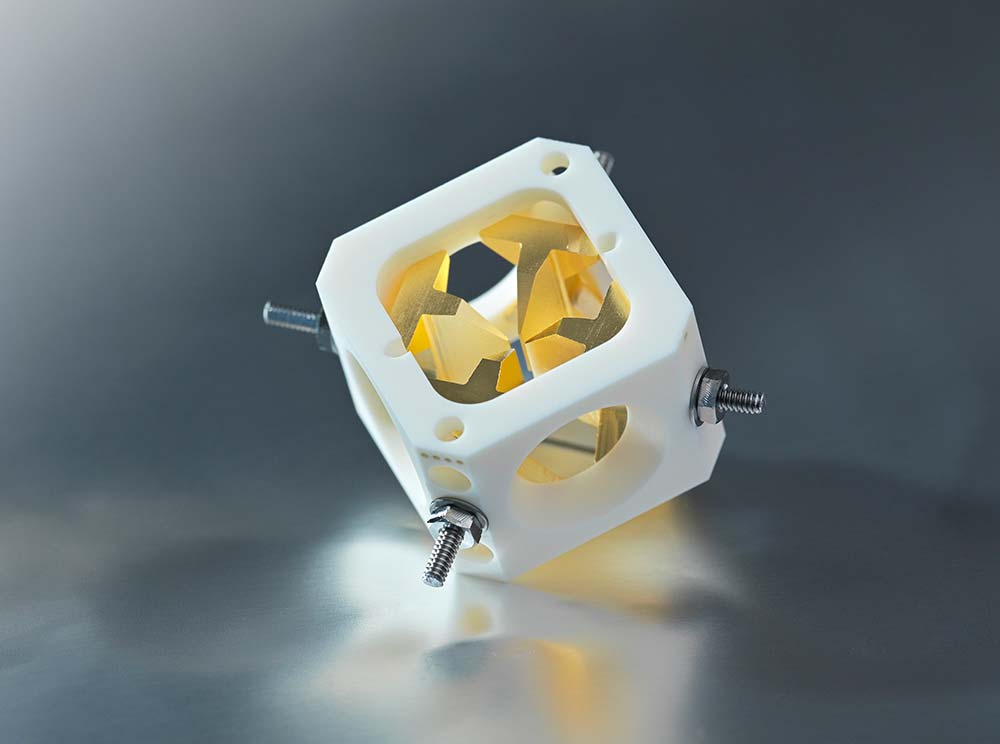}
\hskip3em
\includegraphics[width=.35\linewidth]{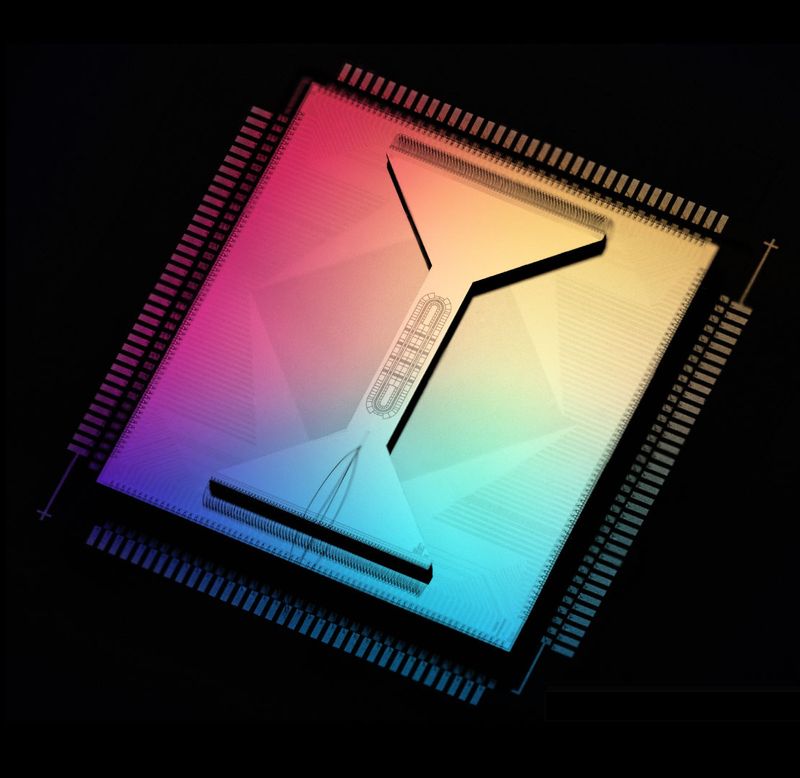}
\caption{Different ion traps technologies.  Left side: a four-rod linear Paul trap.  Source:~\cite{AQTwebsite}.  Right side: a surface-electrode trap.  Source: \cite{quantinuum2023_Moses_Baldwin_Allman_Ancona_Ascarrunz_Barnes_Bartolotta_Bjork_Blanchard_Bohn_etal._2023}.}
\label{fig:trap-flavours}
\end{figure}

The basic idea of the Paul trap is to constrain ions by a so-called \textit{ponderomotive} force that arises from a time-dependent electric field and drives a charged particle towards the minimum of the electric potential which is at the trap centre. This is achieved by a quadrupolar RF field with frequency~$\orf$, which can be created by a variety of electrode configurations. While the original Paul trap was designed for the sole purpose of storing ions, more advanced setups not only can store ions, but also locate them at very distinct positions making them available for precise manipulation.  Two typical examples for different trap types are shown in Fig.~\ref{fig:trap-flavours}.

In the Paul trap, the electric potential is a combination~$\Phi(x,y,z,t) = U(x,y,z) + V(x,y,z,t)$ of a static potential~$U(x,y,z)$ and a dynamic potential~$V(x,y,z,t)$ that oscillates with frequency~$\orf$. Typically, the static potential
\begin{equation*}
    U(x,y,z) = \frac{U_0}2 (2z^2 - (x^2 + y^2))  \qquad (U_0>0)
\end{equation*}
has axial symmetry and confines an ion in \textit{axial direction} ($z$), but not in \textit{radial direction} ($x$, $y$), and the dynamic potential
\begin{equation*}
    V(x,y,t) = \frac{V_0}2 (x^2-y^2) \cos(\orf t) ~,
\end{equation*}
oscillates with frequency~$\orf$ and attracts the ions towards the $z$-axis.

\paragraph{Ion motion in the trap}

In this potential, which is the one of a linear trap, the ion can oscillate in radial or in axial direction.  The motion in axial direction is trivial: the dynamic potential is constant in this direction and therefore, the axial motion is the one of a simple harmonic oscillator.

The radial motion is less trivial. If~$u(t)$ is a radial coordinate of the ion (either~$x$ or~$y$), the oscillation of the ion is described by the \textit{Mathieu equation}
\begin{equation*}
    \frac{d^2}{dt^2} u(t) 
    + \frac{\orf^2}{4}[a + 2q\cos(\orf t)] u(t) = 0 ~.
\end{equation*}
The parameters~$a$ and~$q$ are proportional to~$U_0$ and~$V_0$, respectively.  In the special case~$V_0 = 0$, the time-dependent potential vanishes, implying $q=0$, and the Mathieu equation becomes the differential equation of a harmonic oscillator.  For the given potential, the parameters~$a$ and~$q$ of the Mathieu equation are the same with exception of a sign change of~$q$.

In ion-trap quantum computers, $U_0$ and~$V_0$ are typically chosen such that~$\abs{a}\ll1$ and~$\abs{q}\ll1$.  We also need to assume that~$2\abs a\ll q^2$, because otherwise the motion of the ion becomes unstable.  Under these assumptions, the solution of the Mathieu equation is approximately given by
\begin{equation*}
    u(t)\approx u_0 \cos(\omega_t t)
        \left[1+\frac{q}{2}\cos\left(\orf t\right)\right] ~.
\end{equation*}
Therefore, the motion of the ion is given by a product of two oscillations: a slow \textit{secular oscillation} with the trap frequency~$\omega_t$ and the faster oscillating \textit{micromotion} with frequency~$\orf$ and a much smaller amplitude. The trap frequency~$\omega_{t}$ depends on~$U_0$ and~$V_0$.  Typical frequencies are (see~\cite{Pogorelov_2021}): $\orf \approx 2\pi\cdot\qty{27.4}{MHz}$ and~$\omega_t \approx 2\pi\cdot\qty{3}{MHz}$, so the frequency~$\orf$ of the dynamic potential is an order of magnitude larger than the trap frequency~$\omega_t$.

As we will discuss later, the secular motion is exploited to entangle ions. But the micromotion is an unavoidable artefact and can reduce the reliability of a trapped-ion quantum computer. While this kind of micromotion is minimised when the ions are cooled, another kind, called \textit{excess micromotion}~\cite{Berkeland1998}, can arise when further static electric fields affect the system. This excess micromotion cannot be reduced by cooling because it is completely driven by the electric fields.

\paragraph{Ion chains}

Although so far we only discussed a single ion in a trap, the results are transferable to several ions which can be trapped along the trap axis. If the radial confinement is strong enough, the ions will arrange in a linear chain, called \textit{ion chain} or \textit{ion string}, along the trap axis. The distance between the ions is determined by the equilibrium of the Coulomb repulsion and the potential rise caused by the \textit{endcaps}, the caps at the end of the trap. The endcaps provide the axial confinement, an example for endcaps are the cone-shaped electrodes in Fig.~\ref{fig:trap-scheme}. The typical endcap-to-endcap length is of the orders of millimetres (for instance, Ref.~\cite{Pogorelov_2021} reports a length of~\qty{4.3}{mm}). This leads to typical distances of adjacent ions that are of the order of micrometers (Ref.~\cite{Pogorelov_2021} reports a minimal distance of~\qty{3.4}{\micro m} for a chain of 16 ions with a length of about~\qty{60}{\micro m} while a distance of~\qty{5}{\micro m} is reported in Ref.~\cite{Wang2020High}).

\paragraph{Vibrational modes}

\begin{figure}
    \centering
    \includegraphics[width=0.7\textwidth]{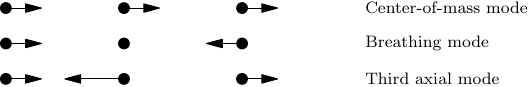}
    \caption{Axial vibrational modes of an ion chain with 3 ions.  The centre-of-mass mode (com) has the lowest frequency~$\omega_\text{com}$.  The breathing mode and the third axial mode have the frequencies~$\sqrt3\,\omega_\text{com}$ and~$\sqrt{29/5}\,\omega_\text{com}$, respectively. Source:~\cite{haffner_2008}.}
    \label{fig:three-axial-modes}
\end{figure}

The oscillatory motion of the ions from their equilibrium position is described in terms of normal modes (vibrational modes). Given a chain of~$N$ ions, there are~$3N$ degrees of freedom for the motion. This translates into~$3N$ eigenmodes of which~$2N$ are radial modes, where the ions oscillate in radial direction, and~$N$ axial modes, where the ions oscillate along the $z$-direction. The more ions in the chain, the denser becomes the excitation spectrum for the vibrational mode.  In general, the frequencies of all these vibrational modes are different, with exception of pairs of radial modes that are degenerate due to the cylindrial symmetry of the potential.  For three ions, for instance, in axial direction, there are three vibrational modes (see Fig.~\ref{fig:three-axial-modes}).
\begin{itemize}
\item \textit{Centre-of-mass mode}: In this mode, the whole chain moves and has a low frequency.
\item \textit{Breathing mode}: In this mode, the ion in the center is fixed and the ions at the chain ends move in contrary direction.  The mode has a medium frequency.
\item \textit{Third mode}: In this mode, all ions oscillate.  The mode has the highest frequency.
\end{itemize}

The larger the number of ions in a chain, the more crowded the oscillation spectrum becomes.  This means that the oscillations are more difficult to distinguish and, at a certain point, they mix, which leads to strong decoherence.

\paragraph{Challenges}

In principle, this trap design is sufficient to perform basic quantum computing operations. The weakness of this approach is the very limited amount of ions that can be reliably controlled. The more ions are present in the trap the more difficult it becomes to individually address them.

This weakness can be overcome by a pseudo-planar version of the linear trap: the surface electrode trap where the geometry of the slabs is unrolled onto a plane and the ions are trapped around~\qty{50}{\micro m} above the electrode plane~\cite{Chiaverini2005, Seidelin2006, Stick2006, Cho2015, Pino_Dreiling_Figgatt_Gaebler_Moses_Allman_Baldwin_Foss-Feig_Hayes_Mayer_et_al._2021, quantinuum2023_Moses_Baldwin_Allman_Ancona_Ascarrunz_Barnes_Bartolotta_Bjork_Blanchard_Bohn_etal._2023}.
This trap design has several advantages: It can be microfabricated and allows to introduce several small regions opening up the possibility to move the ions within the trap. Junctions can be included such that multiple ion arrays can be stored and manipulated~\cite{Blakestad2009}. 

Similar to this design is the sandwich trap, where the electrodes are arranged in two planes and ions are kept between them~\cite{Kaushal2020}. (\grnote{Wouldn't this belong to the surface trap paragraph above or into a new paragraph?}) Another realisation of the surface electrode trap is the high-optical-access trap~\cite{maunz_ionqtrap_2016} that allows to interact with the ions from a wide variety of angles and directions, such as horizontal as well as perpendicular to the trap plane. However, this kind of design is more prone to ion loss than a linear trap due to its smaller potential depth which is of the order of~\qty{100}{meV} (to be compared to the typical trap depths for the linear traps of the order of a few~\unit{eV}). Recently, this problem has been mitigated by a three-dimensional design of a surface electrode trap achieving a potential depth of $\sim\qty{1}{eV}$~\cite{Auchter2022}.

\subsubsection{Ion-light coupling\status{TS,OK}}
\label{sec:ion_light_coupling}

The state of the ions in the trap is given by their respective electronic state ($\ket0$ or~$\ket1$ etc.) and their common motional state, i.e., which vibrational modes are excited and how strong.  Transitions between these states are achieved by monochromatic laser pulses of a particular polarisation.  

In the so-called \textit{Lamb-Dicke} regime, the interaction of the ion with the laser light can be described by a simplified model, in which the transition of the qubit (between~$\ket0$ and~$\ket1$) can be accompanied by a change of the vibrational mode of maximally one phonon. Then, there are three resonances for the laser light (see also Fig.~\ref{fig_ca_levels}a):
\begin{itemize}
\item \textit{Carrier resonance}: the light frequency~$\omega$ is equal to the ion's transition frequency~$\omega_0$.
\item \textit{First red sideband}: the light frequency~$\omega = \omega_0 - \nu$ is equal to the ion's transition frequency~$\omega_0$ minus the vibrational mode's frequency~$\nu$. The excitation of the ion comes with the destruction of a vibration quantum.
\item \textit{First blue sideband}: the light frequency $\omega = \omega_0 + \nu$ is equal to the ion's transition frequency~$\omega_0$ plus the vibrational mode's frequency~$\nu$.  The excitation of the ion comes with the creation of a vibration quantum.
\end{itemize}
In the Lamb-Dicke regime, there are no higher-order sidebands.  Note that light propagating along the axial direction only interacts with the vibrational modes that oscillate in axial direction and light propagating in radial direction only with the likes in radial direction.

In the case of single ions, we still have three different vibrational modes to excite (corresponding to the three space directions).  If these have different frequencies, for instance one axial oscillation
with frequency~$\omega_a$ and two radial oscillations with frequency~$\omega_r$, there are two blue and two red sidebands (see Fig.~9 in~\cite{haffner_2008}).  If we engineer one of these frequencies to be much larger than the other one, we can restrict to one vibrational mode.

\subsubsection{Laser cooling\status{DV,OK}}
\label{sec:cooling}

% explain principles with one example ion, cite others

In a trapped-ion processor, the slow secular oscillation of the ion chain acts as a quantum information bus for entangling operations. To ensure high fidelity in quantum operations, it is crucial to have the ions in a defined pure motional state of the harmonic trapping potential rather than in a thermal mixed state.  After trapping, the kinetic energy of the ions is typically in the region~\qtyrange{0.1}{100}{eV}. Reducing its kinetic energy, or decreasing its velocity, is performed by \textit{laser cooling} and involves applying several laser beams in particular configurations.  This, in general, is realised in two steps: first, by \textit{Doppler cooling} to the Doppler limit and then by employing several sub-Doppler cooling methods, the most common one being \textit{resolved sideband cooling}.

In some cases, other cooling techniques should be employed.  For example, if ions are shuttled, cooling them is a challenge. In those cases, cooling can be achieved by \textit{sympathetic cooling}. This technique requires introducing another ion species in the trap, which is well suited for cooling. Cooling this ion will also cool down the qubit ions in its vicinity. 

In the following, we will limit ourselves to presenting the basic description of these cooling techniques, and refer the reader to the existing literature for more details, e.g.,~\cite{Leibfried2003, Eschner2003, Segal2014,Kielpinski2000, Morigi2001}. 

\paragraph{Doppler cooling}

Doppler cooling~\cite{Hansch1975,Wineland1975} is based on the principle of the Doppler effect and has been extensively used for cooling atoms in the last decades. This technique takes advantage of the fact that when an atom is moving towards a laser beam, for the atom, the frequency of the light is higher than in the reference frame of the laser.  For the atom, the laser frequency then is \textit{blue-shifted} relative to the laser frequency for the laser.  Conversely, when the atom is moving away from the beam, the frequency for the atom is lower and the laser frequency is \textit{red-shifted}.

Let us consider a moving atom interacting with a monochromatic laser beam, and that the laser is \textit{red-detuned}. This means that in its reference frame, the atom has a frequency slightly lower than the resonance frequency of the atom's \textit{cooling transition}. If the atom is moving towards the laser beam, it will see the photons coming from the laser at higher frequencies and therefore closer to resonance. If the atom is moving away from the beam, it will see the photons at even lower frequencies and so further away from resonance.

Being closer to resonance, the photons that move toward the atom will be more likely to be absorbed compared to the photons that are chasing the atom.  As a result of the absorption of the photons moving towards it, the atom also adopts the photon's momentum, and this leads to deceleration. After being absorbed, the atom will spontaneously emit the photon, but this emission occurs isotropic, causing no net change in momentum on average. The result of the deceleration is the progressive cooling (deceleration) of the atom until reaching the Doppler limit.  In a way, Doppler cooling to the ion is like the movement in a viscous liquid.

\begin{figure}
\includegraphics[scale=0.5]{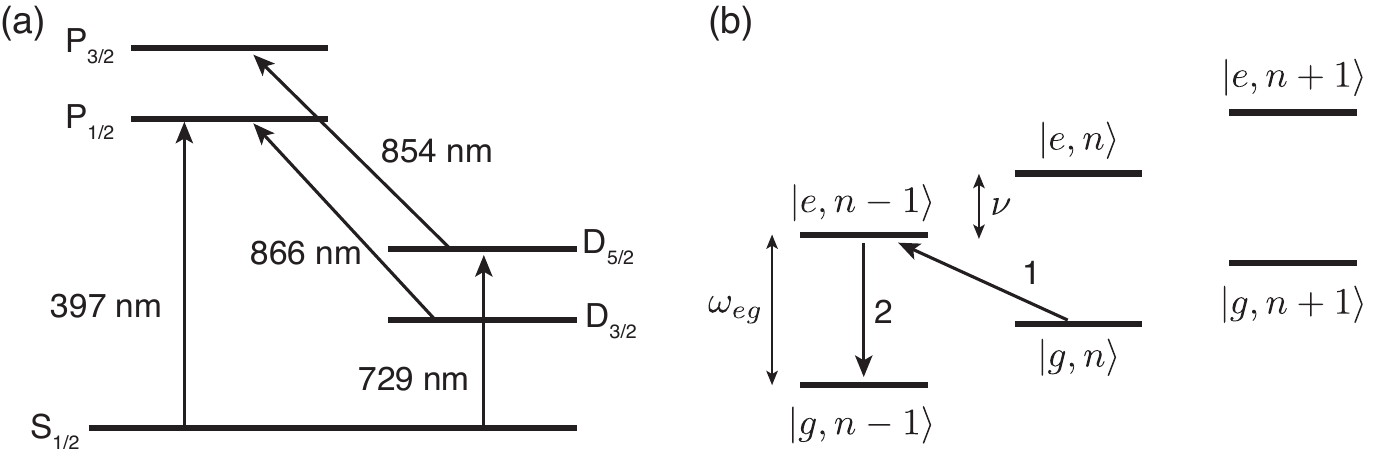}
\caption{(a) Simplified level scheme for the calcium $\isotope[40]{Ca}^+$ ion. (b) General scheme for sideband cooling starting from the state $\ket{g,n} $ via a red sideband $\ket{g,n} \longrightarrow \ket{e,n-1}$ (arrow 1) followed by a spontaneous emission or an additional optical pump process to the state $\ket{g,n-1}$ (arrow 2).}\label{fig_ca_levels}
\end{figure}

The same technique can also be used for cooling trapped ions. When a red-detuned laser beam is directed into the trap, as the ions oscillate, they will move periodically towards the laser and they will absorb the blue-shifted red-detuned photons due to the Doppler effect and consequently slowed down. In this way, their oscillation amplitudes are reduced and the ions are cooled. For example, for the calcium~$\isotope[40]{Ca}^+$ ion, a commonly used ion species, laser cooling is performed on the dipole allowed $S_{1/2}$–$P_{1/2}$ transition by shining a laser at~\qty{397}{nm} wavelength that is red-detuned by half a natural linewidth (see Fig.~\ref{fig_ca_levels}(a) for a simplified energy level diagram). During this process, the ion can end up in the $D_{3/2}$ level that makes the cooling process ineffective. In order to avoid this, another laser (at~\qty{866}{nm} wavelength for the $D_{3/2}$-$P_{1/2}$ transition) is switched on for moving back the ion to the $S$-$P$ cooling cycle.

After Doppler cooling, in a typical ion trap, the state of the ion's motion is described by a mixed state of the vibrational modes of the trap with an average number of~1 to~10 vibrational modes~\cite{Schindler_2013}. Further cooling techniques described below need to be applied in order to cool the ions to the quantum mechanical ground state of the vibrational modes.

\paragraph{Sideband cooling}

After pre-cooling the ions with Doppler cooling, resolved sideband cooling is used to cool ions beyond the Doppler limit and bring them to the vibrational quantum ground state~\cite{Poschinger_2009,Chen_raman_sb_2020}.

Sideband cooling can be explained by considering that the ion's two internal qubit states $\ket g$ and $\ket e$ are coupled with the vibrational motion states $\ket{n}$ ($n=0,1,\dots$) of the trapping potential that can be considered as an harmonic oscillator of frequency $\nu$. We label those states as $\ket{g,n}$ and $\ket{e,n}$ (see Fig.~\ref{fig_ca_levels}(b)). If the trap is strong and the ion is in the Lamb-Dicke regime (see Sec.~\ref{sec:ion_light_coupling}),
%\footnote{In this regime, the Lamb-Dicke parameter~$\eta=z_0 k_z$ is much smaller than one. Here, $z_0$ is the amplitude of the ion's vibration in the motional ground state and~$k_z$ the wave vector of the laser light used for cooling.  The parameter can also be written as~$\eta^2 = \omega_R/\omega_z$, with~$\omega_R$ the recoil energy (typical energy gain if the ion absorbs a photon, and~$\omega_z$ the trap frequency.  This shows that in the Lamb-Dicke regime, the cooling laser cannot excite vibrations of the ion chain.} 
those levels are coupled via ion-laser interactions and can undergo transitions at the carrier frequency and the red and blue sidebands, respectively. 
%three transitions: if the laser frequency is tuned at the energy difference $\omega_{eg}$ between the states $\ket{g,n}$ and  $\ket{e,n}$, the transition (called carrier) couples only the states $\ket{g,n}$ and  $\ket{e,n}$ with no change in the vibrational number $n$. If laser frequency is tuned at $\omega_b = \omega_{eg} + \nu$  the laser will drive the transition $\ket{g,n} \longleftrightarrow \ket{e,n+1}$ called blue-sideband. Finally, if the laser frequency is tuned at $\omega_r = \omega_{eg} - \nu$ the laser will drive the transition $\ket{g,n} \longleftrightarrow \ket{e,n-1}$ called red sideband.

Sideband cooling is obtained by tuning the laser to the red sideband frequency such that the ion undergoes the transition $\ket{g,n} \longrightarrow \ket{e,n-1}$ (arrow 1 in Fig.~\ref{fig_ca_levels}(b)). Then, spontaneous emission via the transition $\ket{e,n-1} \longrightarrow \ket{g,n-1}$ (labeled as arrow 2 in Fig.~\ref{fig_ca_levels}(b)) effectively reduces the mechanical oscillation by one vibrational quantum. This transition can also be helped by actively repumping to $\ket{g,n-1}$ via a third level. After repeating those steps, the state $\ket{g,0}$ is reached and the ion is cooled in the ground state with high probability.

In the case, e.\,g., of the calcium $\isotope[40]{Ca}^+$ ion the red-sideband is the $S_{1/2} \longleftrightarrow D_{5/2}$ transition at~\qty{729}{nm}, while the transition $D_{5/2} \longleftrightarrow P_{3/2}$ at~\qty{854}{nm} wavelength is the transition that helps repumping population from the $D_{5/2}$ state to the $S_{1/2}$.

The procedures outlined in this section are also applicable, after some modification, in the general case where more than one ion is trapped in a linear chain~\cite{Eschner2003}.

\paragraph{Sympathetic cooling}

Sympathetic cooling is another technique used for reducing the kinetic energy of trapped ions. This approach involves using an additional ion referred to as the ``coolant'', which is stored in the same trap as the ion that represents the qubit. By direct laser cooling of only the coolant ion, the qubit ion will also be cooled since these ions share normal modes of motion, because they interact via the Coulomb potential. To prevent the light used for cooling from causing decoherence in the qubit ion, it is crucial that the coolant ion only weakly couples to the internal state of the qubit ion. The interested reader can find the theoretical details behind such cooling technique in \cite{Kielpinski2000, Morigi2001}. Sympathetic cooling of trapped ions has been achieved in various ion combinations such as $\isotope[40]{Ca}^+\mathendash\isotope[40]{Ca}^+$~\cite{Rohde2001}, $\isotope[24]{Mg}^+ \mathendash \isotope[9]{Be}^+$~\cite{Barrett2003},  $\isotope[27]{Al}^+ \mathendash \isotope[9]{Be}^+$~\cite{Rosenband2007}, and $\isotope[171]{Yb}^+\mathendash\isotope[138]{Ba}^+$~\cite{Pino_Dreiling_Figgatt_Gaebler_Moses_Allman_Baldwin_Foss-Feig_Hayes_Mayer_et_al._2021, quantinuum2023_Moses_Baldwin_Allman_Ancona_Ascarrunz_Barnes_Bartolotta_Bjork_Blanchard_Bohn_etal._2023} and recently also for $\isotope[137]{Ba}^+\mathendash\isotope[88]{Sr}^+$~\cite{Delaney_Sletten_Cich_Estey_Fabrikant_Hayes_Hoffman_Hostetter_Langer_Moses_etal._2024}.

Despite being commonly used, sympathetic cooling presents its own set of challenges, in particular regarding the experiment duration and complexity.

Regarding duration, laser cooling of a mixed-species ion chain may take a few milliseconds even for only a single shared mode. Consequently, cooling often dominates the algorithm runtime~\cite{Pino_Dreiling_Figgatt_Gaebler_Moses_Allman_Baldwin_Foss-Feig_Hayes_Mayer_et_al._2021, quantinuum2023_Moses_Baldwin_Allman_Ancona_Ascarrunz_Barnes_Bartolotta_Bjork_Blanchard_Bohn_etal._2023}, and can significantly limit the efficiency of the algorithm and increase the overall experiment time. Then, transporting two-species ion chains can represent a difficult task  when compared to the single-species case. This can hinder the scaling up of quantum computing systems that rely on sympathetic cooling. Some alternatives that show significantly faster performance than the typical duration of sympathetic cooling have been recently proposed in~\cite{fallek2024}.
 
Regarding experiment complexity, trapping a second species requires an additional set of laser sources that greatly increases the complexity of optical elements and can make the design and the implementation of the experimental setup more challenging.

\subsubsection{State initialisation\status{KW,OK}}
\label{sec:state_initialization}

After cooling, the ion needs to be initialised to a specific state, which is the starting point for computations. This is generally done via a technique called \textit{optical pumping}.

The general idea is that the ion is driven by light with a defined polarisation until decaying to a state where the drive becomes ineffective. In order to illustrate the mechanism, we consider the calcium $\isotope[40]{Ca}^+$ ion: After cooling, both the $S_{1/2}$ states with $m=\pm 1/2$ can be populated because either the $P_{1/2}$ or the $P_{3/2}$ states can decay to both the two sublevels of $S_{1/2}$ with the same probability. One can initialise the state to one of the two $m=\pm 1/2$ $S_{1/2}$ states by driving a dipole transition $S_{1/2} \longleftrightarrow P_{1/2}$ with, e.\,g., $\sigma^+$ polarised light. Given the light polarisation, this coupling is effective only between the state $S_{1/2}$ with $m=-1/2$ and the state $P_{1/2}$ with $m=+1/2$. One can drive this transition until the ion eventually decays to $S_{1/2}$ with $m=+1/2$. Once the ion decays to $S_{1/2}$ with $m=+1/2$, then the coupling with the $\sigma^+$ polarised light becomes ineffective, as there is no transition that can be driven by this light, and the ion stays in the $S_{1/2}$ with $m=+1/2$ state.

\subsubsection{Readout\status{DV,TS,OK}}
\label{sec:readout}

\paragraph{State-dependent resonance fluorescence}

A qubit's state is detected by state-dependent resonance fluorescence, implemented using a technique called \emph{electron shelving}~\cite{Dehmelt1975,haffner_2008}.
%an electric-dipole-allowed transition, which is usually identical to the transition used for Doppler cooling is driven and then the presence or absence of ion fluorescence is observed. 
%The basic idea of the electron shelving method is simple. 

Suppose that the qubit states we need to measure are either $\ket{0}$ or $\ket{1}$, and take~$\ket0$ as the ground state and~$\ket1$ as the long-lived excited state of the ion. For electron shelving, we need an additional short-lived excited state $\ket{a}$ (see Fig.~\ref{fig:state-dependent-fluorescence}). By driving the transition $\ket{0} \longleftrightarrow\ket{a}$ with a laser, if the ion is in the state $\ket{0}$, it will decay from the short-lived state $\ket{a}$, scatter the laser photons which can be detected by a photon detector and end up in the state $\ket{0}$. If instead the ion is in the state $\ket{1}$, the detector will register no signal as the states $\ket{0}$ and $\ket{a}$ are not coupled to $\ket{1}$.  Accordingly, the states~$\ket0$ and~$\ket1$ are called the \textit{bright} and \textit{dark state}, respectively.  In the case of a general superposition $\alpha \ket{0} + \beta \ket{1}$, driving of $\ket{0} \longleftrightarrow\ket{a}$  will either cause the scattering of photons with probability $|\alpha|^2$, corresponding to the measurement of $\ket{0}$, or it will generate no scattered photons at all with probability $|\beta|^2$, corresponding to the measurement of $\ket{1}$. In this way, the states $\ket{0}$ and $\ket{1}$ can be distinguished from the scattered light detected by the photon detector.

\iffalse
\begin{figure}
    \centering
    \begin{asy}
        unitsize(1cm);
        defaultpen(fontsize(10pt));
        //--
        real d=0.1; // distance from level line to label
        real l=2;   // length of level line
        void elevel(pair p1, string s) {
            pair p2=p1+(l,0);
            draw(p1--p2,royalblue+linewidth(2bp));
            label(s,p2+(d,0),E);
        };
        pair p0=(1.5,0);
        pair p1=(3,2);
        pair pa=(0,3);
        elevel(p0,"$\ket0$ (bright state)");
        elevel(p1,"$\ket1$ (dark state)");
        elevel(pa,"$\ket a$ (short-lived auxiliary state)");
        draw(p0+(l/2,0)--pa+(l/2,0),Arrows);
    \end{asy}
    \caption{State-dependent fluorescence.  The state~$\alpha\ket0 + \beta\ket1$ is measured in the computational basis~$\{\ket0,\ket1\}$. Illuminating the ion with laser light resonant on the~$\ket0\leftrightarrow\ket a$ transition, with probability~$\abs{\alpha}^2$ yields fluorescence light and projects into~$\ket0$ and with probability~$\abs{\beta}^2$ yields no fluorescence light and projects into~$\ket1$.\tsnote{TS: At the left of this figure, put another figure, regarding optical pumping.}}
    \label{fig:state-dependent-fluorescence}
\end{figure}
\fi

\begin{figure}
    \centering
    \includegraphics[scale=0.8]{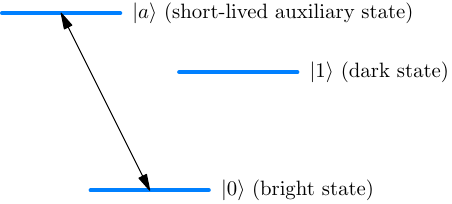}%
    \caption{State-dependent fluorescence.  The state~$\alpha\ket0 + \beta\ket1$ is measured in the computational basis~$\{\ket0,\ket1\}$. Illuminating the ion with laser light resonant on the~$\ket0\leftrightarrow\ket a$ transition, with probability~$\abs{\alpha}^2$ yields fluorescence light and projects into~$\ket0$ and with probability~$\abs{\beta}^2$ yields no fluorescence light and projects into~$\ket1$.\tsnote{TS: At the left of this figure, put another figure, regarding optical pumping.}}
    \label{fig:state-dependent-fluorescence}
\end{figure}

However, this description is subject to some caveats. One reason is that the measurement is not a typical instantaneous projective measurement.  It is usually described in the framework of continuous measurement. In order to end up in an exact projection onto~$\ket{0}$, one needs to measure several times longer than the inverse lifetime of~$\ket{a}$. The reason is that if one performs the projective measurement after a short time, and finds a scattered photon, the state is projected onto $\ket{0}$. However, if the instantaneous measurement finds no photon, there are two possible reasons: the system is in state~$\ket{1}$, which is not resonant with the drive, or the system is in~$\ket{0}$, and there is just not yet a photon there. As consequence, the system is projected into a superposition of $\ket{0}$ and $\ket{1}$.  At the end of the continuous measurement, however, the state will be either $\ket0$ or $\ket1$, consistent with the measurement result.

A typical qubit readout (or measurement) takes between~\qty{300}{\micro s}~\cite{Pogorelov_2021} to~\qty{1}{ms}~\cite{haffner_2008}.

\paragraph{Mid-circuit measurements}

In many quantum algorithms, the qubit register is initialised in the beginning, then the quantum circuit is executed and eventually all the qubits are measured.  In other quantum algorithms, and in particular when implementing quantum error correction, mid-circuit measurements (measurements in the middle of the circuit and not at its end), are needed.  These kind of measurements are problematic because, as explained, a measurement involves the scattering of large numbers of resonant photons.  This leads to a high probability that some of these photons cause errors on nearby qubits and also lead to unwanted excitation of vibrational modes.  Basically, there are two different approaches to achieve mid-circuit measurements with acceptable side effects: (1) moving the qubits that have to be measured away from the others, or into a special area of the ion trap~\cite{Barrett+2004,Chiaverini+2004}; and (2) using a separate ion species, such that the measurement photons are far off-resonant with the other ions' transitions~\cite{Negnevitsky+2018}.

Mid-circuit measurement is a crucial operation for many quantum information protocols.  In particular, it is required for quantum error correction~\cite{Shor1995,Peres1985}, measurement-based quantum computing\cite{Briegel+2009}, teleportation~\cite{Barrett+2004}, entanglement distillation\cite{Bennett+1996} and others.  Another application of mid-circuit measurement is that it allows for the reuse of no longer needed qubits, which in many cases allows for executing quantum circuits with less qubits, or allows to transform quantum circuits into narrower and deeper quantum circuits~\cite{Paler+2016}.

\subsection{Implementing logic gates}
\label{sec:gates}

So far we discussed how to initialise the qubits of a trapped-ion quantum computer. Next, we dive deeper into another crucial part of the computing sequence: the manipulation of the qubits. We describe the implementation of logic gates on single or two qubits.

\subsubsection{Single-qubit gates\status{KW,OK}}
\label{sec:sq-gates}

% Ofxord ionics (Srinivas_etal._2023), NeQxt (Hilder2022), IonQ: wright_benchmarking_2019, Quantinuum (Pino_Dreiling_Figgatt_Gaebler_Moses_Allman_Baldwin_Foss-Feig_Hayes_Mayer_et_al._2021), EleQtron(Bassler2023)

% single SQ gate at a time: Quantinuum (maybe two ions), NeQxt (both do shuttling), Oxford ionics (only test with a single ion), Eleqtron (due to MW approach)
% different SQ gates at different qubits at the same time: IonQ, AQT (with FAOM approach)
% same SQ gate at different qubits at the same time: AQT with AOD approach

A single-qubit gate is a rotation on the Bloch sphere and characterised by the rotation axis and angle.  Such a rotation is achieved by irradiating an ion with light resonant to the qubit transition, having a particular pulse form, phase, electric field strength, and polarisation.

Depending on the type of qubit (see Sec.~\ref{sec:qubits}), the rotation is achieved by different means. For Zeeman and hyperfine qubits, the qubit levels are coupled via a Raman-process, using two co- or counterpropagating laser beams which are detuned from a transition to a higher state, such that their frequency difference matches the qubit transition. Often, these two beams are directed onto a single ion, performing one single-qubit gate at a time~\cite{Pino_Dreiling_Figgatt_Gaebler_Moses_Allman_Baldwin_Foss-Feig_Hayes_Mayer_et_al._2021, Hilder2022, Srinivas_etal._2023}. 
To achieve a parallel execution of single-qubit gates, one of the Raman beams can be sent through a multi-channel acousto-optic modulator (AOM), which allows for individual control of laser amplitude and phase for each beam at its output~\cite{wright_benchmarking_2019}. Thus, different singe-qubit gates can be applied to different ions simultaneously. The second Raman beam is not changed and applied globally to all ions.  
As mentioned in Sec.~\ref{sec:qubits}, another approach for controlling hyperfine qubits employs a global microwave pulse combined with a magnetic gradient~\cite{Bassler2023}. Due to the latter, the transition frequency of each qubit is different, allowing for addressing of single qubits. Here, only a single-qubit gate is applied to one ion at a time. 

For optical qubits, the levels can be directly coupled using laser light. There are different addressing approaches, as described in~\cite{Pogorelov_2021}, which allow for parallelisation of single-qubit gates. In the first variant, the light is send through a splitting module with a fixed number of output channels. The separated beams are then individually modulated by fiber-coupled AOMs and thus, different types of single-qubit gates can be executed on different ions in parallel. The second variant uses a single acousto-optical deflector (AOD) that splits the incoming laser beam. Driving the AOD with multiple RF tones can be used to implement the same single-qubit gate at multiple ions, coming at the cost of creating additional beams below and above the ion chain. 

If the $z$-axis is chosen to align with the light beam's direction, only rotations about an axis perpendicular to the $z$-axis can be implemented. The actual rotation axis is determined by the choice of the light's polarisation vector, the phase or, e.g., in the case of Raman beams, by the phase of the microwave beat note of the laser beams\cite{Pino_Dreiling_Figgatt_Gaebler_Moses_Allman_Baldwin_Foss-Feig_Hayes_Mayer_et_al._2021, Hilder2022, Srinivas_etal._2023}. 
In this way, $x$- and $y$-rotations can be performed, but $z$-rotations are not directly possible. Nevertheless, they can be decomposed, for instance, into a sequence of an $x$- and two $y$-rotations.\footnote{There are also methods to implement a $z$-rotation directly, which uses an off-resonant pulse and the AC-Stark effect.}  The rotation angle is equal to the \textit{pulse area} and, in the simple case of a square pulse, it is given by~$\theta = \Evec\dvec \tau/\hbar$, where~$\Evec$ is the amplitude of the electric field, $\dvec$ is the transition's dipole moment, and~$\tau$ is the duration of the pulse.  We will use the notation~$R_x(\varphi)$ and~$R_y(\varphi)$ for a rotation about the $x$ and $y$-axis, respectively, with an angle~$\varphi$.  The Pauli gates then are~$\qgate{X}\equiv R_y(\pi)$ and~$\qgate{Y}\equiv R_y(\pi)$.

\subsubsection{Two-qubit gates\status{DV,OK}}
\label{sec:tq-gates}

%See:~\cite{calcium_trap} (p.~62)

Two-qubit gates act on a pair of qubits and enable the creation of entanglement among them. Combined with single qubit rotations, entangling two-qubit gates form a universal gate set of operations.  Some commonly used two-qubit entangling gates include the controlled-NOT (\qgate{CNOT}) gate, which flips the target qubit if the control qubit is in the state $\ket{1}$, and the controlled-\qgate{Z} (\qgate{CZ}) gate, which applies a phase shift $\qgate{Z}$ to the target qubit if the control qubit is in the state~$\ket{1}$.

There are several strategies for implementing two-qubit gates.  The ``classical'' two-qubit gate is the Cirac-Zoller gate~\cite{cirac_zoller_1995_PhysRevLett.74.4091}, which is a~\qgate{CZ} gate.  Its implementation requires only single ion addressing and ground-state cooling of the vibrational mode.

The general idea of the Cirac-Zoller gate is to use the vibrational mode of the ion chain as a buffer for mediating the interactions among the qubits and modify their internal states. Basically, a laser pulse is directed at the first ion, transferring its internal excited state amplitude to a vibrational mode. This vibration is shared with the entire ion chain and
can acquire  a phase conditional on the state of the second qubit. Finally, the modified state of the vibrational mode is transferred back to the first qubit. This enables the realisation of the~\qgate{CZ} operation.
%For an explicit description of this gate, see Appendix~\ref{app:tq-gates}.

The disadvantage of this gate is that it only works if the vibrational mode of the ion chain is cooled down to its ground state beforehand. This is very challenging and the reason why the  Cirac-Zoller gate is basically not used in trapped-ion quantum computing. The broadly used alternative is the M\o{}lmer-S\o{}rensen gate (MS gate)~\cite{molmer_sorensen_1999}, which does not require ground state cooling of the ion chain. The implementation of the MS gate requires a bichromatic laser field that irradiates the ion chain. The two tones of the laser are symmetrically detuned from a red and blue sideband such that single-photon processes are suppressed. On the contrary, by choosing carefully the frequencies of the two tones, the two-photon processes interfere in a way that makes the internal state dynamics insensitive to the vibrational state while coupling and creating entanglement among the qubit states.

\subsection{Decoherence and noise}

\subsubsection{Qubit lifetime, coherence time and gate fidelities\status{TS,OK}}
\label{sec:lifetime-coherence-time}

A real noisy quantum computer behaves considerably different from an ideal quantum computer, the device that we usually have in mind when we develop quantum algorithms or circuits.  Qubits, even when they are not operated, lose their information when time passes.  The reason is that the interaction of a qubit with its environment, including other qubits, cannot be completely avoided.  Both, the qubit \textit{lifetime} ($T_1$-time) and its \textit{coherence time} ($T_2$-time) are important parameters that characterise this information loss of single qubits.
%Refer to App.~\ref{app:lifetime-coherence-time} for a more detailed explanation of lifetime and coherence time.

When the qubits are operated on, as part of a quantum gate, there are further sources of imperfections.  The frequency of the control laser may be slightly different from the ion's qubit transition frequency.  And the length of the control pulse could be too large or too small.  Both cause non-ideal quantum gates: rotations about slightly skewed axes or with slightly too long or short rotation angles.  Imperfections of quantum gates are not measured with the $T_1$- and $T_2$-parameters but with the concept of fidelity. Suppose that starting with states $\ket{\psi_1^{\mathrm{in}}}, ..., \ket{\psi_N^{\mathrm{in}}}$, a perfect $N \times N$ gate would yield the desired target states $\ket{\psi_1^{\mathrm{trgt}}}, ..., \ket{\psi_N^{\mathrm{trgt}}}$ but the real noisy implementation of the gate yields $\ket{\psi_1^{\mathrm{out}}}, ..., \ket{\psi_N^{\mathrm{out}}}$. Then, in principle, the gate fidelity is a measure of the difference between the desired and actual output states.

Maintaining long qubit lifetimes and coherence times and high gate, state preparation and measurement fidelities is a primary challenge in quantum computing. Factors such as environmental noise, electromagnetic fluctuations, and even residual motion of the trapped ions can impact these times.

\subsubsection{Sources and types of noise\status{Karen,OK}}
\label{sec:decoherence}

Usually, the lifetimes of the internal energy states of trapped ions are quite long, i.e., in the order of seconds or more (see Tab.~\ref{tab:Parameters_1}) but the motional states in the trap can easily be perturbed, which is called \textit{motional heating}. Thereby, the slow secular motion of ions in the trap is excited and the motional quantum number is increased, leading to reduced coherence times and two-qubit gate fidelities. Even for the MS gate, which is independent of the motional state, changes in the motional quantum number can lead to dephasing~\cite{Webb2018}. %\tsnote{Why? Didn't we say that MS gate works also with excited motional modes?}

The main source for motional heating is electric field noise.  Even when the ions are cooled to the quantum ground state, some energy typically remains, resulting in the ions having a non-zero electric dipole moment, which in turn makes them susceptible to very small electric field fluctuations~\cite{Brownnutt_2015}. The motional heating strongly depends on the distance between the ions and the trap and is related to various processes at the trap surface, of which not all are fully understood on a microscopic level. 

Nevertheless, different types and sources of electric field noise are known in the context of ion traps~\cite{Lakhmanskiy_2019}. (1) \textit{Technical noise} originates from lab devices such as RF drive electronics or DC power supplies and thus does not depend on the trap frequency or chip temperature. It can be reduced, e.\,g., by using low noise electronics and electronic filtering of all connection lines that enter the vacuum cell. (2) Another fundamental source of noise is \textit{Nyquist noise} or thermal noise, which originates from thermal fluctuations of charge carriers, being present in any resistor. It is often difficult to distinguish from technical noise, but can be mitigated by using special materials with low electrical resistance for the trap electrodes and connection lines. (3) In addition, there is \textit{surface noise} originating from metallic and dielectric surfaces near the ion such as the trap chip itself, which can have a significant influence. Cleaning of the trap surface as well as cooling the setup to cryogenic temperatures (about 10K) can greatly reduce the level of surface noise. 

Further sources of decoherence are magnetic field noise, fluctuations of laser intensity, phase and frequency, off-resonant excitation of ions and collisions with background gas~\cite{Brandl_2016}. The latter can be reduced by lowering the pressure in the vacuum cell, which is also facilitated by cooling the setup. Off-resonant light shifts and photon scattering can degrade quantum operations, and let ions become trapped in undesirable internal states. To overcome this, additional repumping lasers are needed to reinitialise the ion to the~$\ket{0}$ state, adding complexity to the setup.

In general, spatially-varying external perturbations such as magnetic field inhomogeneity or Stark shifts can lead to de facto differences between the otherwise identical qubits.

\subsubsection{Losing ions and recovery\status{DV,OK}}
\label{sec:ion-loss}

Apart of the effects of noise discussed in the previous section, trapped-ion quantum computers suffer from two further error types: ion loss and leakage~\cite{Stricker2020}.

\textit{Ion loss} refers to the physical disappearance of an ion from the trap.  Loss typically occurs due to interactions of the trapped ions with background gas molecules.  The loss rate increases with the number of ions and the residual gas pressure in the ion vacuum chamber.  In trapped-ion systems with a large number of ions needed for useful computations (several thousands) and single-ion lifetimes in the order of tens of hours, the loss of an ion is expected to occur at least once in a second.  Loss is a detrimental limitation for scaling.

Detecting a lost ion is straightforward but reloading it and correcting the error in the computation is much more challenging.  Nevertheless, there are reliable methods to reload ions without disturbing the others~\cite{bruzewicz2016} and codes to cope for the caused error in the computation~\cite{Vala_2005,kang2023,connolly2023}.

\textit{Leakage} is due to the fact that an ion is not a two-level system, as a qubit.  Leakage occurs when the computational space of a single qubit in an ion is not well isolated, allowing for unintended transitions from the qubit's state to other states of the ion.  This can be caused by imperfections in the control and manipulation of qubits, such as imperfect or not very well calibrated laser beams. 

Leakage can be reduced by designing appropriate correction protocols, for example, by swapping or teleporting qubits to the computational subspace~\cite{Aliferis2007}. Additionally, in topological quantum memories where the logical information is stored in ensemble of physical qubits, to prevent leakage, error correction schemes have been developed e.g. in~\cite{Fowler2013}. Correction for generalised leakage has been also demonstrated experimentally for minimal instances of, e.g., the surface code in~\cite{Stricker2020}. There leakage is first detected by coupling the ion with an ancilla qubit and reading the state of the ancilla via a quantum non demolition measurement. After the detection, a code-switching protocol can be implemented to restore the logical information on the remaining physical qubits.

\section{Challenges of the trapped-ion platform}
\label{sec:platform_overview.tex}

The goal of this section is to give an overview of hardware-related properties of the trapped-ion platform, discuss their implications on scaling, quantum error correction, as well as applications, and hint to current possibilities for accessing different trapped-ion quantum computers.

\subsection{Scalability challenges\status{TS,OK}}
\label{sec:scalability}

We discussed the advantages of trapped-ion quantum computers, in particular the very high single-qubit and two-qubit fidelities, the high initialisation and measurement fidelities and the all-to-all connectivity.  The high fidelities allow for quantum error correction and the all-to-all connectivity allow for two-qubit gates not only between next neighbours but also between two distant qubits and removes the need of many \qgate{SWAP} operations.

But there are also disadvantages like the slow gate speed, which implies long computation times and the scaling of the trapped-ion quantum computer up to many qubits.  Scaling is inevitable in order to get into the regime, where quantum computing becomes useful for industrial applications.

To illustrate the scaling challenges, we imagine a ``standard system'', a linear trap with several tens of ions. Realising such a system is still feasible with reasonable effort.  Imagine now that we want to scale up such a quantum computer to many thousands of qubits. Then, several severe challenges appear (see, e.\,g., \cite{Kaushal2020,schwerdt+2024-scalable}).

\paragraph{Physics challenges}

The most important scaling challenges imposed by physics are listed in the following.
\begin{itemize}
\item The \textit{heating rate}: With a larger number~$N$ of ions, the heating rate increases drastically. The electric field noise has a high influence, especially on the low-frequency modes.  In particular, the axial center-of-mass mode, whose frequency typically decreases as~$1/N$, is affected.  The heating rate of this mode can increase linearly with~$N$~\cite{Brownnutt_2015}.
\item The \textit{spectral crowding}: The larger the ion chain, the more dense becomes the energy spectrum of the normal modes. Then, resolving the modes and avoiding crosstalk between them becomes more difficult and the gate times of entangling gates grow rapidly with~$N$ (for radial modes possibly as~$N^2$~\cite{shapira+2023-fast}), and therefore the computations become slow.
\item The \textit{distance between two adjacent ions} becomes smaller with higher~$N$ and therefore, it becomes more difficult to address one single ion with a laser beam.
%
% 2112.10655
\item The \textit{collision rate of background gas molecules} in the vacuum chamber with the ions increases with the number of ions in the chain. This limits the possible depths of quantum circuits or requires complicated methods to reconstruct the ion chain after an ion got lost (see Sec.~\ref{sec:ion-loss}).
\item The ion string approaches an \textit{instability}, which causes it to deform into a zig-zag arrangement~\cite{Landa+2013}.  To avoid this, with increasing number of ions, the ratio between the radial and the axial trap strength has to become larger.
\end{itemize}

\paragraph{Engineering challenges}

Apart of these challenges related to the physics of trapped-ion quantum computers, there are also severe engineering challenges, that also grow considerably with the number of ions. In particular, cramming optical components for laser cooling, repumping, fluorescence detection and addressing each ion individually for state manipulation into a small volume, given by the distance between the ions, is very difficult. 

Scaling trapped-ion systems for quantum computing also requires optimising control electronics. A key challenge is managing RF power dissipation in the oscillating electric fields that confine ions. As trap arrays grow for more qubits, RF heating increases, limiting scalability. However, a recent study~\cite{Sterk_etal._2024} presents a surface ion trap confining up to 200 ions while reducing RF power dissipation. By elevating electrodes and removing dielectric material, their design minimises this critical factor restricting larger-scale trapped-ion systems. Further electronics advances can enable truly scalable arrays.

\subsection{Remedies to the scalability issues\status{Intro text: TS,OK}}

%\tsnote{Add the idea of 2-qubit gates based on Rydberg atoms.  See, e.\,g., \url{https://www.nature.com/articles/s41586-020-2152-9} and \url{https://journals.aps.org/prx/pdf/10.1103/PhysRevX.7.021038}.}

There are different concepts to improve the scaling of trapped-ion quantum computers.  If on the order of 100 ions is sufficient, the linear Paul trap can be kept and the long ion string is divided into different substrings. This is described in Sec.~\ref{sec:divide-string}.  For intermediate size trapped-ion quantum computers with on the order of 1,000 or 10,000 ions, the linear Paul trap is no longer appropriate and traps with different zones are needed, see Secs.~\ref{sec:traps-with-different-zones} and~\ref{sec:qccd-arch}.  Eventually, if even larger systems are needed, various trapped-ion quantum computer chips have to be connected.  This can be achieved with directly stitching chips together or by using photonic interconnects.  We outline this in Sec.~\ref{sec:photonic-interconnects}.

\tsnote{We did not explicitly discuss the technology of Universal Quantum! Should do that.}

\subsubsection{Localized phonon modes\status{TE,TS,OK}}
\label{sec:divide-string}

If the number of ions in the linear trap gets to the order of 100, the scalability challenges in Sec.~\ref{sec:scalability} become pressing.  A way out is the concept demonstrated in~\cite{Olsacher2020}.  The idea is to use a linear Paul trap but to localise vibrational modes by dividing a long ion chain into different zones containing an ion subchain each.  This is achieved through pinning individual ions with optical tweezers.  Each of these subchains have their own vibrational modes and modes of different subchains do not interact.  Therefore, in parallel, in each of these zones a gate can be performed.  The optical tweezers are programmable, giving a large amount of flexibility.  This concept cures many scaling issues but it is no improvement to the situation when a two-qubit gate with very distant ions has to be performed.

\subsubsection{Traps with different separate zones and shuttling\status{TS,OK}}
\label{sec:traps-with-different-zones}

For in the order of 1,000 or 10,000 ions, a linear trap is no longer sufficient.  In this case traps with several different \textit{zones} are used, which hold the ions.  Gates between two ions from different zones are realised by bringing the ions together in some \textit{interaction region}.  Moving an ion around in the trap is called \textit{shuttling} or \textit{ion transport}.  While moving the ion solves the immediate problem, it also incurs a large execution time overhead.  Traps with different zones are usually micro-fabricated surface trapped-ion chips.  An example is the QCCD (quantum charge-coupled device) architecture (see Sec.~\ref{sec:qccd-arch}). Here, however, due to ion shuttling and ion cooling, the computation is also slow. The impact of shuttling is further discussed in Sec.~\ref{sec:impact-of-shuttling}.  Traps with different zones allow to optimize the different zones, for instance to have one zone where fast gates can be performed (calculation qubits) and another one where the lifetimes are as large as possible (memory qubits).  A further challenge is that when one of two entangled qubits is moved, the highly susceptible state will probably suffer particularly.

Other alternatives are to dynamically split the long ion chains with thousands of ions into much smaller segments, by introducing large spaces between them or fixing some ions, both by using optical potentials (optical tweezers).  It is also possible to use many traps each of which features a manageable amount of ions and photonic interconnects between them or to move ions between these traps. Both alternatives cause a considerable cost if entanglement between segments or traps needs to be created or manipulated~\cite{schwerdt+2024-scalable}.  

Generally, one can say that trapped-ion quantum computing faces two key scaling challenges: the "wiring problem" of providing individual control signals to each qubit, and the "sorting problem" of moving qubits to enable connectivity. While there is no optimal solution to the scaling challenge and all solutions have their advantages and disadvantages, there are further ideas to improve different aspects of the trapped-ion quantum computing for better scalability.

\iffalse
% section does not fit here
\subsubsection{Integrated local control via forced motion [ThE,\TBD]}

One approach introduces a new paradigm for qubit control by leveraging the electric fields used to trap and move ions around the chip, allowing a single high-frequency structure to control large numbers of qubits in parallel. It uses the trapping electrodes to apply small oscillating voltages that make the ions vibrate, changing how they respond to a shared high-frequency qubit-control field, providing localised control of each qubit. By reusing the transport electrodes for qubit control and integrating the low-frequency voltages using existing microelectronics industry techniques, this forced-motion addressing achieves full parallel control with minimal added complexity \cite{Srinivas_etal._2023}.

\fi

\subsubsection{QCCD architecture\status{TE,TS,OK}}
\label{sec:qccd-arch}

A typical architecture for trapped ions is the \textit{QCCD (quantum charge-coupled device) architecture}, which was proposed as a scalable method for trapped-ion quantum computation~\cite{Wineland_Monroe_Itano_Leibfried_King_Meekhof_1997, Kielpinski_Monroe_Wineland_2002, Pino_Dreiling_Figgatt_Gaebler_Moses_Allman_Baldwin_Foss-Feig_Hayes_Mayer_et_al._2021} and is motivated by the considerations in the previous section. For a recent review of potential implementations and its challenges see \cite{Akhtar_Bonus_Lebrun-Gallagher_Johnson_Siegele-Brown_Hong_Hile_Kulmiya_Weidt_Hensinger_2023, Kaushal2020}. 

The QCCD architecture's primary objective is to realise a scalable and highly accurate trapped-ion quantum computer. This ambition implicates several considerable prerequisites, as described in  \cite{Pino_Dreiling_Figgatt_Gaebler_Moses_Allman_Baldwin_Foss-Feig_Hayes_Mayer_et_al._2021}.
\begin{itemize}
    \item The apparatus must possess the capability to confine numerous short ion chains or just separate ions, each able to execute high-precision operations.
    \item It necessitates fast ion transfer for efficient movement between these ion chains.
    \item The system mandates meticulous monitoring of qubit phases, alongside precise synchronisation of control signals across disparate regions.
    \item There is a potential need to capture two distinct ion species~-- one functioning as a qubit, while the other serves to sympathetically cool the ions to near-motionless states post-transport.
    \item The architecture must facilitate the parallelisation of transport and quantum operations across the device.
\end{itemize}
We discuss further solutions to these prerequisites in Sec.~\ref{subsec:further}.

While some solutions have been proposed for some of these challenges (see~\cite{Labaziewicz_2008, maunz_ionqtrap_2016, Bowler_Gaebler_Lin_Tan_Hanneke_Jost_Home_Leibfried_Wineland_2012,Kaushal_Lekitsch_Stahl_Hilder_Pijn_Schmiegelow_Bermudez_Müller_Schmidt-Kaler_Poschinger_2019,Barrett_DeMarco_2003}), combining all of these features into one machine creates complex performance requirements. For example, high-quality qubit operations need very small disturbances from motion. This means electrode voltages must have very low noise levels. But fast transport requires high-bandwidth voltage control for the same electrodes. Previous work, like \cite{Wan_Kienzler_Erickson_Mayer_Tan_Wu_Vasconcelos_Glancy_Knill_Wineland_et_al._2019, Kaufmann_Ruster_Schmiegelow_Luda_Kaushal_Schulz_von_Lindenfels_Schmidt-Kaler_Poschinger_2017, Home_McDonnell_Szwer_Keitch_Lucas_Stacey_Steane_2009}, has made good progress in making QCCD quantum computers more scalable. However, these efforts either lacked multiple zones for parallel operations or sympathetic cooling, or were limited to just one qubit pair.

A relatively recent architecture, which allows scaling, is the so-called \textit{race track architecture}, where the trap is organised in a closed loop of traps in the form of a race track
%. This was first implemented  the case of Quantinuum H2 processor. 
\cite{quantinuum2023_Moses_Baldwin_Allman_Ancona_Ascarrunz_Barnes_Bartolotta_Bjork_Blanchard_Bohn_etal._2023}.
% moved here from Sec. 4
%\subsection{QRAM at the horizon[ThE - ready]}
This architecture uses a separation into computational (or ``active'') and ``parking'' zones.  In the former, the gates between qubits are performed.  
The latter ones can be thought of a kind of memory region where qubits can be stored more safely until they are involved in a gate operation or be embedded into a network mode for also storing entanglement~\cite{Drmota2023}.  ``Parking'' zones are optimised for long coherence times.

A very recent demonstration that solves both the wiring and the sorting problem employs a grid-based surface electrode Paul trap~\cite{Delaney_Sletten_Cich_Estey_Fabrikant_Hayes_Hoffman_Hostetter_Langer_Moses_etal._2024}, which is a 2D chip design that minimises the number of control signals needed, using a fixed number of analog signals plus one digital input per qubit, while also enabling efficient qubit sorting through its grid arrangement. The study demonstrates qubit transport and sorting with a (physical) swap rate of 2.5 kHz and very low heating, indicating the quality of the control system. The (physical) swap rate demonstrates the potential of a 2D grid layout, because it is much quicker to rearrange qubits on a grid in contrast to qubits in a 1D setting (line or loop).

Another ansatz to solve the wiring problem in QCCD architectures has been suggested in~\cite{Malinowski_Allcock_Ballance_2023}. It avoids the common approach of having one or multiple control lines for each qubit, which makes scaling to large qubit numbers challenging. Instead, 
%the so-called WISE (Wiring using Integrated Switching Electronics) architecture 
it integrates switching electronics into the trap chip in a way that allows to address/control large qubit numbers with fewer wires.

%\kwnote{Moved from Sec. 5.4, check and integrate }
Overall, the advantage of the QCCD architecture is its ability to maintain low error rates even for complex circuits comprising many transport and gate operations. It achieves the maximal quantum volume \dbnote{Do we have a Ref. for this claim?} and demonstrates small crosstalk.

\subsubsection{Photonic interconnects: photon-mediated entanglement\status{TS,OK}}
\label{sec:photonic-interconnects}

We have seen in Sec.~\ref{sec:qccd-arch} that the QCCD architecture with its different trapping zones on one chip addresses several of the scaling challenges.  Shuttling, however, causes a considerable overhead and together with the cooling effort limits the number of ions on a QCCD.  Building even larger systems then involves multiple separate QCCDs, that are linked via photonic interconnects.  Two ions in separate QCCDs then can be entangled by causing them to emit a photon each, funneling these photons into an optical fiber and interfering them with a beam splitter.  This is called photon-mediated entanglement. 

The key difficulty here is the very low collection efficiency for the emitted photons.  Experiments have used large lenses with a high aperture to get an acceptable (but low) collection efficiency~\cite{carter2023ion} or, in on-chip architectures sophisticated gratings~\cite{knollmann2024integrated}.  Photon-mediated entanglement is possible, but it also incurs a large overhead and the entanglement of a larger number of ions from one QCCD with a larger number of ions from another one will arguably be prohibitive.

There have been successful experiments entangling two single trapped ions via photonic links, realizing quantum key distribution~\cite{Nadlinger2021} or to demonstrate a network of optimal clocks~\cite{Nichol2021}. Moreover, trapped ions haven been employed as memory qubits connected to a photonic detection system via a quantum link based on optical fibers, realizing so-called blind quantum computing~\cite{Drmota2023}. The idea behind this is that a client can execute a secret quantum computation on a remote device which content is also not known to the device supplier.

\subsection{Further considerations on scalability}
\label{subsec:further}

%One concept for quantum information processing used in neutral atoms is leveraging the Rydberg interactions. There are also attempts (not yet commercially available) to use a short-range Rydberg dipole-dipole interaction and a fast $\qty{700}{\nano s}$ entangling gate between trapped ions, using $\isotope[88]{Sr}^+$ and two-photon excitation to couple the $\ket{0}$ and $\ket{r}$ state transition \cite{Zhang_Pokorny_Li_Higgins_Pöschl_Lesanovsky_Hennrich_2020, Higgins_Li_Pokorny_Zhang_Kress_Maier_Haag_Bodart_Lesanovsky_Hennrich_2017}.  It should be noted that the reported fidelity to produce a Bell state is 78\%, and a total error of less than 0.2 per cent is predicted for experimentally achievable parameters. 

\subsubsection{Ion shuttling and its impact\status{TS,OK}}
\label{sec:impact-of-shuttling}

For traps with different zones, two-qubit gates on two ions in different zones cannot be directly implemented.  The solution is to bring one of the ions next to the other, by moving it through the trap(s).  This operation, however, is relatively slow and impacts the overall circuit execution time.  In addition, moving an ion always increases its decoherence, which causes a reduction of the gate fidelity.  The use of efficient shuttling techniques is, therefore, vital for improving the overall performance of quantum gates. Different shuttling strategies and methods have been developed and used to minimise the shuttling impact on circuit execution time, while still ensuring high gate fidelity.  Efficient ion shuttling techniques will remain critical for achieving practical scalability and high-fidelity quantum gate operations.

The impact of shuttling on the execution time can be quantified. Table~1 in~\cite{quantinuum2023_Moses_Baldwin_Allman_Ancona_Ascarrunz_Barnes_Bartolotta_Bjork_Blanchard_Bohn_etal._2023} reports a typical circuit time budget of 1\% vs.\ 58\% vs.\ 41\% for quantum operations vs.\ ion transport vs.\ cooling.  According to~\cite{Walther_Ziesel_Ruster_Dawkins_Ott_Hettrich_Singer_Schmidt-Kaler_Poschinger_2012}, moving an ion over a distance of~\qty{200}{\micro m} takes about~\qty{10}{\micro s}.

On the basis of such numbers, the system size and execution time for a quantum processor that factorises a 2048 bit number using the Shor algorithm has been estimated\cite{Lekitsch2017}. They assume a single qubit gate time of~\qty{2.5}{\micro s}, a two-qubit gate time of~\qty{10}{\micro s}, an ion separation and shuttling time of~\qty{15}{\micro s} each, a static magnetic field gradient ramp-up and ramp-down time of~\qty{5}{\micro s} each, and a measurement time of~\qty{25}{\micro s}.  This results in a total error correction cycle time of~\qty{235}{\micro s}.  The conclusion is that in this model, factorising a 2048-bit number takes on the order of 110 days and requires a system size of~$2\cdot 10^9$ trapped ions.
%\cite{Kaushal2019}

\subsubsection{Qubit reuse\status{TE,OK}}
\label{sec:qubit_reuse}

The efficient use of available qubits for computation in quantum computers is crucial due to the current limited number of qubits. Qubit reuse, which involves resetting and reusing qubits after a mid-circuit measurement (see Sec.~\ref{sec:readout}), is a promising method to improve the efficiency. An automated framework was developed for compiling quantum circuits that incorporates qubit reuse, compressing circuits effectively~\cite{DeCross_Chertkov_Kohagen_Foss-Feig_2022}. While qubit-reuse compilation shows promise, the trade-offs between reduced qubit count, increased circuit depth, and associated error implications must be carefully considered. Recent research has focused on practical benefits, and an 80-qubit MaxCut quantum approximate optimisation algorithm (QAOA) problem has been successfully solved on a 20-qubit quantum processor using qubit-reuse compilation algorithms~\cite{DeCross_Chertkov_Kohagen_Foss-Feig_2022}.

\subsubsection{Qudits\status{TS,DB,OK}}
\label{sec:qudits}

As pointed out previously, due to the limited number of qubits in trapped-ion quantum processors, the available qubits have to be used as efficiently as possible. In the following, we mention another method to use the limited quantum hardware resources.

Quantum computers usually rely on two-state systems to encode information via qubits. However, many quantum systems that serve as basis for qubits --- like ions in the present case, see Fig.~\ref{fig:term-scheme-group-II-ions}, --- are intrinsically multi- or infinite-dimensional. This poses the question: why, as a unit of quantum information, do we use qubits? Instead of encoding information into binary systems, we could also encode information into systems with~$d$ possible values, introducing the \textit{qudit} as its quantum information carrier. These systems offer new possibilities to leverage coherence and entanglement while being more error resilient, making them quite interesting for future technological advancements.

An implementation of qudits in a trapped-ion quantum computer has already been demonstrated~\cite{RingbauerEtAl2022}. The study uses $\isotope[40]{Ca}^+$-ions and the levels~${}^2S_{1/2}$ and~${}^2D_{5/2}$. In a static magnetic field, the six degenerate Zeeman sublevels of~${}^2D_{5/2}$ and the two sublevels of~${}^2S_{1/2}$ split up and allow for a qudit with eight levels. The selection rules allow for ten transitions and a rich set of quantum gates for the qudits can be implemented. While it is possible to adapt the M\o{}lmer-S\o{}rensen gate for qudits, also native qudit gates can be constructed~\cite{Hrmo:2023}.

With qudits available, $d$-dimensional quantum systems could be simulated natively without decomposition into a binary representation, or qubit-based algorithms could be implemented more efficiently.

\subsection{Fault-tolerance and error correction\status{DV,OK}}
\label{sec:ft-qec}

%\tsnote{Explain state-of-the-art and current developments in qec on the trapped-ion platform.  For material, see p.~33 of the Supplemental Information of \url{https://www.nature.com/articles/s41586-022-05434-1}.}

Due to the high precision of gates and control, trapped-ion quantum computers have already been used to produce and manipulate logical qubits~\cite{da_Silva_Ryan-Anderson_Bello-Rivas_Chernoguzov_Dreiling_Foltz_Gaebler_Gatterman_Hayes_Hewitt_etal.}, which is directly paving the way for near-term fault-tolerant quantum computing on physical hardware. For a general overview of quantum error correction and fault-tolerant circuit design, we refer the reader to our review paper on neutral atoms~\cite{Wintersperger2023} and various in-depth introductions to the topic, such as~\cite{Gottesman_2010}.

The successful execution of fault-tolerant operation in a quantum processor relies heavily on a series of critical steps: the initialisation of logical states according to a quantum error correcting code; the measurement of the error patterns (syndrome extraction) and the subsequent error correction; the realisation of universal logical gate sets; and the measurement of the logical quantum states. All these operations need to be carried out in a fault-tolerant way in order not to introduce errors that cannot be tolerated by the underlying quantum error correcting code. 

In recent years, there have been numerous experimental achievements in almost all the previous steps in trapped-ion quantum processors. These experiments have shown the potential of trapped-ion systems in addressing and mitigating errors within quantum computations. In this discussion, we present the most recent and significant experimental advances.

The state preparation of error correcting codes has been achieved, e.g., in~\cite{nigg2014}, where a single logical qubit is encoded via the Steane code~\cite{Steane_1996} using seven trapped-ion qubits. This encoding allows the detection of a single bit flip, a single phase flip, or a combination of both, regardless of which qubit the errors occur on. Recently, in~\cite{Wang_Simsek_Gatterman_Gerber_Gilmore_Gresh_Hewitt_Horst_Matheny_Mengle_etal._2023} a three dimensional $\left \llbracket 8,\, 3,\, 2 \right \rrbracket$ colour code has been implemented with eight trapped-ion qubits. This code allows a transversal non-Clifford $\qgate{CCZ}$ gate and has been used to realise a fault-tolerant one-bit adder circuit.

Errors in quantum codes are detected via error syndrome readout by measuring the code stabilisers (parity operators) on the logical qubits. Different approaches have been developed for realising  fault-tolerant syndrome measurements, e.g.~in~\cite{linke2017} where the two stabilisers of a four-qubit error detection code are measured fault-tolerantly or in~\cite{Hilder2022}, where the stabilisers of the Steane code are measured fault-tolerantly by using flag qubits that detect hook errors, i.e., faults occurring on the syndromes that spread onto the data qubits.

Quantum error correction cycles that are necessary to protect the logical states from the faults that might happen have been realised for the first time in~\cite{schindler2011} in a three-ion system for a phase-flip or recently in~\cite{ryananderson2021} for the seven-qubit Steane code capable of correcting both phase and bit flip errors.

Lastly, control and measurement of logical qubits have been also recently demonstrated. For example, in~\cite{egan2021} 13 trapped-ion qubits are encoded in a logical state that is then rotated at logical level. In~\cite{Erhard2020} operations on two logical qubits are implemented via the so-called lattice surgery, that consists of merging and splitting groups of physical qubits arranged on lattices, while in~\cite{ryananderson2021}  logical $\qgate{CNOT}$ gates are performed with five-qubit and seven-qubit codes.

More involved logical operations (e.g., fault-tolerant preparation of logical magic states or fault-tolerant logical $\qgate{T}$ gates) have been achieved with the Steane code in~\cite{Postler2021}.

\section{Hardware state-of-the-art\status{FD,OK}}
\label{subsec:explanation_criteria}

The set of relevant hardware properties in our recent paper on neutral-atom quantum computers~\cite{Wintersperger2023} was selected under the condition that it must be general enough to characterise many, though not necessarily all, quantum computing platforms. Since neutral-atom and trapped-ion quantum computers have similar building blocks, we reuse this set for the discussion of the trapped-ion platform. In the following, we will discuss these parameters and provide typical values for the trapped-ion platform in Tab.~\ref{tab:Parameters_1}. In general, the properties of the underlying quantum hardware influence the quality and runtime of the computation in many ways. This influence can be direct, such as when considering operation fidelities, or more indirect, such as the connectivity between qubits. An introduction to the interplay between hardware properties and algorithm performance can be found in~\cite{Wintersperger2023}.   

\begin{table*}[htbp]
    \centering  
    \begin{tabularx}{\linewidth}{p{5cm}X}
       \bf{Parameter}  &\bf{Typical values today (near future)}\\
       \hline
       \hline
       \noalign{\vskip 1mm}    
       \multicolumn{2}{l}{\bf{Qubits}}\\
        Amount & $\sim 32-36$~\cite{Quantinuum_System_Model_H2_V1.2, ionq_forte_data_2024} %\kwnote{IonQ Forte: 36}~\cite{ionq_forte_data_2024} \slnote{Quantinuum H2: 32~\cite{Quantinuum_System_Model_H2_V1.2}, NeQxt: n.a., eleQtron: n.a.} 
        ($\sim 1000$~\cite{Malinowski_Allcock_Ballance_2023})  \\
        Connectivity & all-to-all~\cite{egan_2021,Quantinuum_System_Model_H2_V1.2, ionq_forte_data_2024}\\
        Multiple states (i.e., qudit) &  In principle possible\cite{Hrmo:2023, Ringbauer2022}\\
        \noalign{\vskip 0.5mm} 
        \hline
        \noalign{\vskip 1mm} 
        \multicolumn{2}{l}{\bf{Lifetimes and decoherence times}}\\
        Trap lifetime & $\sim$ minutes to days~\cite{bruzewicz2016, Zhang2017, Bruzewicz2019}\\
	Decoherence times & $T_1\sim$\qtyrange[range-units=single,range-phrase=--]{10}{100}{s}~\cite{ionq_forte_data_2024}, $T_2 \sim$\qty{1}{s}~\cite{ionq_forte_data_2024, AQT_Marmot}\\ 
		\noalign{\vskip 0.5mm} 
        \hline
        \noalign{\vskip 1mm} 
        \multicolumn{2}{l}{\bf{Native gates}}\\
        single qubit gates & RXY~\cite{egan_2021}, virtual RZ~\cite{egan_2021},  GPi, GPi2,  U${}_{1,q}$, RZ \tsnote{Wrong ref?} \\
        Two-qubit gates: & XX (via M\o{}lmer-S\o{}renson-interaction)~\cite{egan_2021}, ZZ, M\o{}lmer-S\o{}renson-Gate, RZZ \\
       Parallelism & Segmented traps / QCCD: parallel application of gates in different gate zones possible~\cite{Pino_Dreiling_Figgatt_Gaebler_Moses_Allman_Baldwin_Foss-Feig_Hayes_Mayer_et_al._2021}. Linear traps: parallel execution of 2-qubit gates possible by optical segmentation using optical tweezers~\cite{Olsacher2020} \\
        \noalign{\vskip 0.5mm} 
        \hline
        \noalign{\vskip 1mm} 
        \multicolumn{2}{l}{\bf{Fidelities of operations}}\\
        1-qubit gate  &\numrange[range-phrase=--]{0.9996}{0.999999}~\cite{AQT_Marmot, ionq_forte_data_2024, Quantinuum_System_Model_H2_V1.2, Harty2014}\\ %\kwnote{IonQ Forte: 0.9998} \slnote{Quantinuum H2: 0.9998-0.99999~\cite{Quantinuum_System_Model_H2_V1.2}, AQT Marmot \cite{AQT_Marmot}: 0.9996, NeQxt: n.a., eleQtron: n.a.}; (0.999999 achieved~\cite{Harty2014}) \\
        2-qubit gate  & \numrange[range-phrase=--]{0.985}{0.9982}~\cite{AQT_Marmot, ionq_forte_data_2024,Quantinuum_System_Model_H2_V1.2}\\ %\kwnote{IonQ Forte: 0.996} \slnote{Quantinuum H2: 0.995-0.9982~\cite{Quantinuum_System_Model_H2_V1.2}, AQT Marmot: 0.985, NeQxt: n.a., eleQtron: n.a.} \\ 
         Readout &  0.995 (single-shot detection fidelity)~\cite{egan_2021}\\%, 0.998 \fdnote{We only have a single number available which is a combined measure for Readout and Preparation error} \\
	    % Preparation &  0.998 \fdnote{see Readout} \\ 
		SPAM	& \numrange[range-phrase=--]{0.995}{0.999}~\cite{ionq_forte_data_2024, Quantinuum_System_Model_H2_V1.2}\\ %\kwnote{IonQ Forte: 0.995} \slnote{Quantinuum H2: 0.995-0.999~\cite{Quantinuum_System_Model_H2_V1.2}, NeQxt: n.a., eleQtron: n.a.}  \\	
        \noalign{\vskip 0.5mm} 
        \hline
        \noalign{\vskip 1mm} 
        \multicolumn{2}{l}{\bf{Execution times}}\\
        1-qubit gate & \qtyrange[range-units=single,range-phrase=--]{1}{110}{\micro s}~\cite{Malinowski_Allcock_Ballance_2023, Srinivas_etal._2023, Leu2023, Pogorelov_2021, Saner2023} (\qty{1}{\micro s}~\cite{Malinowski_Allcock_Ballance_2023}), (\qty{110}{\micro s}~\cite{Saner2023})
        \\
        2-qubit gate & \qtyrange[range-units=single,range-phrase=--]{15}{900}{\micro s}~\cite{Malinowski_Allcock_Ballance_2023, Pogorelov_2021, Saner2023} (\qty{100}{\micro s}~\cite{Malinowski_Allcock_Ballance_2023})\\ %\kwnote{no ref found for \qty{15}{\micro s}, \cite{Clark2021} is about Bell state preparation and anyways demonstrates \qty{35}{\micro s}} \dbnote{Ref.~\cite{Saner2023} shows a M\o{}lmer-S\o{}rensen gate in \qty{15}{\micro s}}\\
        Ion shuttling & \dbnote{Do we still want to put a number?} \\ 
        Preparation & \dbnote{Do we have a number here?} \\ 
        Readout & \qtyrange[range-units=single,range-phrase=--]{300}{1000}{\micro s}~\cite{Pogorelov_2021, haffner_2008} \\% \dbnote{Added values from Sec.~\ref{sec:readout}} \\
        %Average depth-1 circuit time (proposed benchmark) & 28 ms (only one value reported) \cite{Quantinuum_System_Model_H2} \\ %\fdnote{We only have a single number available. Should we skip it?} 
        %Single gate layer time & (\qtyrange[range-units=single,range-phrase=--]{0.4}{25}{ms}~\cite{Malinowski_Allcock_Ballance_2023}) \dbnote{Addition to other metric?} \\
        \noalign{\vskip 0.5mm} 
        \hline
        \noalign{\vskip 1mm} 
        \multicolumn{2}{l}{\bf{Installation and operation}}\\
        Required infrastructure & Vacuum cell and pumps, lasers, optical elements, microwave sources, signal generators and modulators, magnetic field coils. \\
        Calibration & No calibration of individual qubits necessary, but regular calibration of controlling devices required; reloading of ions due to possible loss\\
        Specificity & Shuttling operations, Phonon bus \\
        Access & Via cloud offered by the quantum computing provider or public cloud services~\cite{azure_quantum_2023, google_marketplace_2023, aws_braket_2023} \\
        \hline
        \hline
    \end{tabularx}
    \vspace{0.1cm}
    \caption{Summary of parameters for trapped ion quantum computers, representing the current state-of-the-art with an extrapolation to the future given in brackets. The values reported here reflect the status of the devices run by trapped-ion quantum computing companies, with the perspective of being used for applications already now or in the near future. A description of each parameter as well as references for their values can be found in Sec.~\ref{subsec:explanation_criteria}.}
    Note the trade-offs: a given qubit could be optimised for execution time, or for fidelity, or even for both. Only optimising, e.\,g., for execution time is obviously less challenging than optimising for both.
    \label{tab:Parameters_1}
\end{table*}				

\subsection{Qubits and their connectivity\status{DB,OK}}
\label{subsubsec:qubits_and_connectivity}

Trapped-ion quantum computers utilise cooled and individually addressable ions as qubits, which are typically confined using electromagnetic fields in a trap, such as a Paul~\cite{Raizen1992} or Penning trap~\cite{Sameed:2020}. \textit{Qubit connectivity} refers to the number of qubits with which one qubit can directly interact as well as to how far away those qubits are from each other within the hardware platform. Therefore, it defines a key metric that determines a platform's capability to perform entangling operations. In trapped-ion quantum computers, entangling operations can usually be applied between arbitrary pairs of qubits, providing full, i.e., all-to-all, connectivity. This yields maximal flexibility when generating or optimising quantum circuits, as no~\qgate{SWAP} gates are necessary. Another feature, mostly present in small ion chains, is the ability to entangle more than two, or even all ions in the chain, i.e., performing $N$-qubit gates with $N > 2$.

\subsection{Native gates\status{OK}}
\label{subsubsec:native-gates}

A quantum gate abstracts a unitary operation acting on one or more qubits.  It is considered \textit{native} for a certain quantum computer, if its action on the qubits can be directly realised on the underlying quantum hardware without the need to decompose it into other gates.

A \textit{universal set of quantum gates} comprises a small set of single- and two-qubit gates, with which all unitaries can be approximated.  A typical universal gate set is formed by the gates~\qgate{CNOT}, \qgate{H} and~\qgate{T}.  The primary objective of forming a universal native gate set for a particular platform is to identify a set of gates that is well-suited for the hardware and minimises the total number of gates required to approximate arbitrary unitary transformations.

A quantum computer capable of approximating any unitary transformation must implement a universal native gate set. Table~\ref{tab:Parameters_1} presents a collection of all native gates found in various trapped-ion quantum computers. This set represents a combination of different universal gate sets, indicating that there are multiple methods for achieving universality.

What are the reasons for this variety of gates? On the one hand, there are technical reasons related to the ability to efficiently execute a gate given the properties of the trapped ions and the available controllable interactions. Here, efficiency encompasses qualities such as accuracy and speed. On the other hand, it is sometimes advantageous to have additional gates beyond the universal set available, as it can increase the efficiency of the computation. Let us consider the list of native gates, that we found for the trapped-ion platform, as shown in Tab.~\ref{tab:Parameters_1}.

On a single-qubit level, arbitrary local gates are natively available, which means any kind of rotation of the state vector in the single-qubit Hilbert subspace can be implemented as a native operation.  Note the remark on $z$-rotation in Sec.~\ref{sec:sq-gates}.

Additionally, note the gates $\qgate{GPi}(\phi)$, $\qgate{GPi2}(\phi)$ and $\qgate{U}_{1,q}$~\cite{Quantinuum_System_Model_H2}\footnote{\qgate{GPi} is a $\pi$-rotation and~\qgate{GPi2} a $\pi/2$-rotation, both about the same axis $\nvec=(\cos\phi, \sin\phi, 0)$.}, which are designed with a focus on flexibility. The~\qgate{GPi} gate is parameterised with a phase angle~$\phi$. It flips the computational basis states $\ket{0}$ and $\ket{1}$ while attaching the phases $e^{i\phi}$ and $e^{-i\phi}$, respectively. Thus, it can realise not only the standard gates~\qgate{X} for $\phi=0$ and~\qgate{Y} for $\phi=\pi/2$, but also any arbitrary combination $\qgate{X}\cos\phi+\qgate{Y}\sin\phi$. Similarly, \qgate{GPi2} consumes a phase~$\phi$ as parameter and puts the computational basis states in a superposition while adding a relative phase $e^{\pm i\phi}$ in between them, depending on the input state.
%This corresponds to $R_X(\pi/2)$ or $R_Y(\pi/2)$ for $\phi=\pi$ or $\phi=\pi/2$, respectively.
By using~\qgate{GPi} and~\qgate{GPi2}, every rotational operation on a single qubit can be performed (a $z$-rotation can be decomposed into a sequence of $y$-, $x$- and $y$-rotation). The $\qgate{U}_{1,q}$ gate demonstrates its flexibility by combining all $R_x$ and $R_y$ rotations into a single native gate. Finally, the~\qgate{VirtualZ} gate exemplifies efficiency in design. It models an $R_z$ gate by applying the corresponding phase shift to all subsequent operations, eliminating the need to initiate a separate interaction with the qubits and wait for it to complete.

% NOTE: GPi, GPi2 sind von IonQ and U$_{1,q}$ von Quantinuum

On a two-qubit level, the gates are usually implemented by a M\o{}lmer-S\o{}rensen-type interaction, as already discussed in Sec.~\ref{sec:tq-gates}. When examining the list of two-qubit gates in Tab.~\ref{tab:Parameters_1} you might notice that the most prominent two-qubit gate, the~\qgate{CNOT} gate, is not native to any trapped-ion quantum computer we studied. Instead we found the M\o{}lmer-S\o{}rensen gate and the $R_{xx}$, $R_{yy}$, and $R_{zz}$ rotations. The most flexible among them is the M\o{}lmer-S\o{}rensen gate. It puts the incoming state into a superposition of itself and its bit-flipped counterpart while also introducing a relative phase difference dependent on the incoming state and input parameters. This enables the native realisation of an~\qgate{XX}, \qgate{YY}, \qgate{YX}, or~\qgate{XY} gate. Finally, the $R_{zz}$ rotation offers the possibility to solely add an input-state-dependent phase, with the~\qgate{ZZ} gate as a special case of it.

We conclude the discussion of native gates by drawing attention to the~\qgate{SWAP} gate. It is not included in the list in Tab.~\ref{tab:Parameters_1}, either because it is implicitly native, as QCCD architecture-based systems, where ions are shuttled around to achieve all-to-all connectivity, or because it is unnecessary in systems that achieve all-to-all connectivity with a phonon bus.

\subsection{Fidelities of operations\status{FD,OK}}

Fidelity is a key metric that quantifies how closely the action of a real quantum gate  aligns with the perfect quantum gate that it aims to implement. It is also used to specify the difference between perfect state preparation and measurement (SPAM) and its actual implementation. Fidelity is represented by a number between 0 and 1, with 1 representing perfection. However, achieving high fidelity is challenging due to noise and decoherence, leading to ongoing efforts to improve hardware, error correction techniques, and error mitigation strategies for reliable quantum computations and applications.  For more details refer to Sec.~\ref{sec:lifetime-coherence-time}.

We have collected three common types of fidelities and summarised them in Tab.~\ref{tab:Parameters_1}. The fidelities of single-qubit gate markedly surpass 0.9995, indicating a highly reliable operation compared to the two-qubit fidelities, which range from 0.996 to 0.999. The SPAM fidelity is lower than the single-qubit fidelities and falls within the range of the two-qubit fidelities.

\subsection{Execution times\status{FD,OK}}
\label{subsubsec:execution-times}

We have previously touched upon the topic of execution times when discussing native gate sets in Sec.~\ref{subsubsec:native-gates}. Now, we take a closer look. A complete quantum computation typically consists of three subsequent steps: state preparation, gate execution, and measurement.\footnote{Mid-circuit measurements are disregarded here.} The success of the computation depends on its total execution time being shorter than the coherence time of the qubits. Therefore, reducing execution times is crucial. To provide insight into this aspect, we have compiled a summary in Tab.~\ref{tab:Parameters_1} for the trapped-ion platform.

These numbers enable us to explore various interesting questions and aspects. They provide insights into the overall computational speed, offering a quick estimate of the number of operations that can be successfully executed, disregarding gate errors and focusing solely on decoherence. In this case, the importance of different time components depends on the specific circuit being used. If mid-circuit measurements are not utilised, then measurement and state preparation occur only once, bringing the spotlight onto gate execution times. However, once error correction is introduced, mid-circuit measurement and state preparation operations also become significant contributors to the overall execution time.

In any case, the goal is to execute the operations as fast as possible without sacrificing accuracy. Since gate execution times on the trapped-ion platform are relatively slow, researchers are exploring new elements to create faster native gate sets~\cite{Bazavan2022, Saner2023}.

Another important aspect to consider regarding the overall execution time of a circuit is the capability to achieve \textit{parallel gate execution}. This implies that different gates acting on different qubits can be applied in parallel if the gate sequence permits it. However, achieving this requires even more precise control.

Long ion chain devices, which rely on a single ion chain, must avoid interference with the densely packed spectrum of vibrational modes when addressing the different qubits for parallel gate operations (see also Sec.~\ref{sec:scalability}). This requires highly optimised signals for reliable control~\cite{Figgatt2018}. Another approach is the effective optical segmentation of the ions using optical tweezers~\cite{Olsacher2020} which depends on engineering the phonon spectrum to enable parallel entangling gate operations can be performed reliably. This approach also offers the possibility to implement multi-qubit gates.

Devices utilizing a QCCD architecture, which employ different computation zones, execute gates in parallel by relocating the corresponding ions into well-separated zones~\cite{Srinivas_etal._2023,quantinuum2023_Moses_Baldwin_Allman_Ancona_Ascarrunz_Barnes_Bartolotta_Bjork_Blanchard_Bohn_etal._2023}. The trade-off is the time required for the ion shuttling (see also Sec.~\ref{sec:impact-of-shuttling}). Reducing this time entails increasing the shuttle velocity, which in turn decreases the fidelity of the qubits realised by the ion due to heating and possibility of ion loss.

\section{Overview of applications realised with trapped-ion quantum computers\status{TE,KW,OK}}
\label{subsec:application_overview.tex}

%\ahnote{Thinking loudly: what is specific for ion trap QC? Arent't these applications "the best qualified" e.g. for superconducting QC?}
%\tenote{the benefit for natural qubits (ions and neutral atoms and photons) is their intrinsic qubit stability / fidelity. for neutral atoms and ions (in QCCD) also  the (in principle) all to all connectivity (OK with physical shuttling, not logical (and thus error introducing) SWAP gates), the superconducting cannot do that, they are man made and thus different and difficult to calibrate, have quick decoherence and do not physically move around.}

In general, the potential value that is added by using a quantum computer~-- or a quantum computing component~-- for solving a task depends highly on the problem at hand. In the era of NISQ devices, however, also the fit between the task and the specific properties of the employed quantum hardware platform play a role.

In comparison to other quantum computing hardware platforms, trapped-ion quantum computers have several advantages such as considerably high gate fidelities and long coherence times.  Moreover, two-qubit gates can be applied between any pair of qubits, leading in general to shallower circuits as there is no need for~\qgate{SWAP} gates. Therefore, trapped-ion quantum computers have been used for several types of applications, ranging from quantum chemistry over optimisation to quantum machine learning, all of which are also relevant in industrial settings~\cite{QUTAC_Bayerstadler_2021, QUATC_Awasthi_2023, QUTAC_Riofrío_2023}.
Moreover, simulations of fundamental physics problems have also been carried out. In the following, we describe different applications which have already been implemented on trapped-ion quantum computers.

We remark that according to our knowledge, the following companies now offer access to their quantum computers for other companies or researchers: AQT, IonQ, Oxford Ionics, Quantinuum.  Furthermore, eleQtron, neQxt, and Universal Quantum will do so soon.

\subsection{Industrial applications of trapped-ion quantum computers\status{TE,KW,OK}}
\label{subsec:industry_applications}

\subsubsection{Quantum chemistry}

Certain approaches evaluate the potential of using trapped-ion quantum computers in material simulation or quantum chemistry, and show the substantial impact of optimal circuit synthesis and the benefits of all-to-all connectivity and high-fidelity operations in trapped-ion quantum devices. A common near-term algorithm for quantum chemistry is the variational quantum eigensolver (VQE), which has successfully been implemented on trapped-ion quantum computers to estimate the ground states of simple chemical compounds such as the hydrogen molecule, lithium hydride (LiH)~\cite{Hempel2018}, or the water molecule~\cite{nam_ground-state_2020}.

In~\cite{Goings_Zhao_Jakowski_Morris_Pooser_2023}, the boundaries of VQE are pushed by designing the circuit specifically for trapped-ion quantum computers, using the benzene molecule as a specific example. Since the current quantum computers (including trapped-ion devices) usually exhibit a very limited number of qubits, approaches to decompose circuits into smaller pieces are being developed, as in~\cite{kawashima_optimizing_2021}: The electronic structure problem for a ring of~10 hydrogen atoms, in an approximation requiring~20 qubits, is solved using only~10 qubits by decomposing the circuit into two 10-qubit circuits.

\subsubsection{Combinatorial optimization}

The high quality of the qubits and their control in trapped-ion devices makes them also promising candidates to run near-term algorithms such as the quantum approximate optimization algorithm (QAOA). In~\cite{Obst_Barzen_Beisel_Leymann_Salm_Truger_2023}, the performance of QAOA for the MaxCut problem on trapped-ion and superconducting qubits is compared, showing the suitability of trapped ions for this algorithm. In~\cite{Zhu_Zhang_Sundar_Green_Huerta_Alderete_Nguyen_Hazzard_Linke_2023}, the authors demonstrate on a trapped-ion quantum computer that QAOA results improve with the number of rounds for multiple problems on arbitrary graphs of up to~12 qubits, despite the presence of hardware noise. 
Moreover, reusing qubits can help to solve larger problems on quantum devices with less qubits, which is in particular interesting for ion traps and has been successfully applied to solve an 80-qubit MaxCut problem with QAOA on a 20-qubit trapped-ion quantum computer~\cite{DeCross_Chertkov_Kohagen_Foss-Feig_2022}.

\subsubsection{Quantum machine learning}

Another application of current trapped-ion quantum computers is quantum machine learning (QML).  Especially in the field of generative models, quantum versions of the neural networks can lead to an exponential advantage, such as the learning of joint probability distributions using copulas, which can be directly mapped to multipartite maximally entangled states~\cite{Zhu_2022}.  In~\cite{Zhu_2022}, a quantum generative adversarial network and a quantum circuit Born machine are being used for said task.
In\cite{zhu_copula-based_2023}, these methods are then applied to model risk aggregation of three- and four-dimensional datasets.  Another application of a quantum-classical generative algorithm that has been implemented on trapped-ion devices is the generation of handwritten digits~\cite{Rudolph_2022}.

\subsection{Application of trapped ions as simulators in physics research\status{TE,KW,OK}}

Trapped-ion quantum computers have also been employed to simulate the behaviour of various physical systems.  This includes the exploration of exotic quantum phases of matter with true long-range order:
the inherent long-range interactions of a chain of trapped ions enable the study of so-called continuous-symmetry breaking (CSB) phases, demonstrating the potential of trapped-ion quantum computers to contribute to the understanding of complex quantum phenomena, using it as a simulator~\cite{Feng_Katz_Haack_Maghrebi_Gorshkov_Gong_Cetina_Monroe_2023}.
Another application of trapped-ion devices for fundamental physics is the direct simulation of polarised-light-induced electron transfer~\cite{sun_quantum_2023}, which is relevant for many processes in chemistry, biochemistry and energy-science.  In this work, three energy levels of the ions have been used, realising \textit{qutrits}, a special case of the qudits discussed in Sec.~\ref{sec:qudits}, leading to a higher efficiency of the simulation. A further interesting application is the study of non-equilibrium phase
transitions on a quantum computer~\cite{Chertkov_Cheng_Potter_Gopalakrishnan_Gatterman_Gerber_Gilmore_Gresh_Hall_Hankin_etal._2023}. Here the authors study a quantum circuit that they call the one-dimensional Floquet quantum contact process, which is a quantum circuit generalisation of the contact process, a simple model for disease spreading by contact between infected individuals. In~\cite{Chertkov_Cheng_Potter_Gopalakrishnan_Gatterman_Gerber_Gilmore_Gresh_Hall_Hankin_etal._2023}, the authors demonstrate qubit re-use allowing them to simulate 73 qubits with 20 physical qubits.
Moreover, the realisation of non-Abelian anyons has been demonstrated using trapped-ion quantum computers~\cite{Iqbal_Tantivasadakarn_Verresen_Campbell_Dreiling_Figgatt_Gaebler_Johansen_Mills_Moses_et_al._2024}.

\section{Discussion\status{TE,OK}}
\label{sec:discussion}

In the preceding sections, we highlighted crucial attributes of trapped-ion quantum computers and their impact on the performance of quantum algorithms and the implementation of error correction schemes.  Like any contemporary quantum computing hardware, trapped ions possess both, advantages and drawbacks.  This section delves into a discussion, drawing comparisons with other established platforms like superconducting qubits or neutral atoms.
%Section Sec.~\ref{subsec:disc_applications} focuses on applications where ion trap quantum computers have proven successful and areas where this platform might excel. To wrap up our review, we offer a conclusive summary. We keep in mind that we address applications for industry relevant applications  \cite{QUTAC_Bayerstadler_Becquin_Binder_Botter_Ehm_Ehmer_Erdmann_Gaus_Harbach_etal._2021} and also for fundamental research.

\subsection{Advantages\status{KW,OK}}
%%% advantages

A fundamental advantage of trapped ions is that they are so-called \textit{natural qubits}, as all ions of the same species are identical and have the same properties.  Thus, there is no need for calibration of each single qubit as e.\,g., for superconducting circuits.
As mentioned before, very important features of trapped ions are the high fidelities for gate operations and state detection, reaching two-qubit gate fidelities of~$99.99\%$ (see Tab.~\ref{tab:Parameters_1}) as well as the long lifetimes and coherence times unmatched by any other quantum computing platform.

Moreover, trapped-ion quantum computers can perform two-qubit gates between arbitrary pairs of qubits, rendering an effective all-to-all connectivity.  This is either realised via the common vibrational mode in linear traps or by moving qubits for a two-qubit gate next to each other (shuttling) as in the QCCD approach which is described in Sec.~\ref{sec:qccd-arch}. Shuttling takes additional time, but bringing two qubits together to perform a gate usually results in a higher gate fidelity than introducing~\qgate{SWAP} operations which furthermore need to be decomposed into native gates.
The arbitrary connectivity renders the insertion of costly~\qgate{SWAP} gates unnecessary and therefore leads to shallower circuits and allows for running algorithms already on current NISQ devices in a meaningful way (see Sec.~\ref{subsec:industry_applications}).
In linear ion traps, more than two qubits can be entangled in one operation, corresponding to the direct implementation of gates with three or more qubits.  This avoids the decomposition of these higher-order interactions into several two-qubit gates and thus reduces the circuit depth.  Moreover, as mentioned in Sec.~\ref{sec:qudits}, the natural occurrence of more than two discrete energy levels in atoms and ions can be used to realise qudits, which can reduce the number of ions needed for a computation. 

The QCCD architecture allows to move ions into compute zones with differently optimised properties, e.\,g., fast gate application or maintaining long lifetimes, which also eases the implementation of fault-tolerant quantum computation. 

In general, trapped-ion quantum computers can operate at room temperature.  Often, cooling with a cryostat is used to reduce noise (see Sec.~\ref{sec:decoherence}), but the required temperatures are higher than, e.\,g., for superconducting qubits, offering a more cost-effective performance.

In a very recent work~\cite{da_Silva_Ryan-Anderson_Bello-Rivas_Chernoguzov_Dreiling_Foltz_Gaebler_Gatterman_Hayes_Hewitt_etal.}, researchers show results which signify an important transition from noisy intermediate-scale quantum computing to reliable quantum computing, and demonstrate advanced capabilities toward large-scale fault-tolerant quantum computing.

\subsection{Disadvantages and ways to address them\status{TE,KW,OK}}

%%% disadvantages
% slow gate speed
% scaling to higher number of qubits: QCCD/shuttling, connecting several traps with photonic interconnects 
% shuttling: takes additional time + might induce decoherence

Despite these advantages, trapped-ion quantum computers currently also face challenges, one major aspect being the scaling to qubit numbers larger than~$\sim 50$.  Possible remedies (see Sec.~\ref{sec:scalability}) are to either connect several linear traps by physically shuttling the ions in the two-dimensional QCCD architecture or by separate traps and photonic interconnects.  Moreover, the speeds for qubit manipulation and readout are relatively low, as well as the speed for physically shuttling the ions through the computational zones.  We recall that Table~1 in~\cite{quantinuum2023_Moses_Baldwin_Allman_Ancona_Ascarrunz_Barnes_Bartolotta_Bjork_Blanchard_Bohn_etal._2023} reports a typical circuit time budget of 1\% vs.\ 58\% vs.\ 41\% for quantum operations vs.\ ion transport vs.\ cooling.  Also here, like with most architectures, massive scaling (order of~1000 and more qubits) is a challenge in this case because using shuttling over large distances implies paying a high price in runtime.

Although all qubits within a quantum system share identical properties, discrepancies in the errors associated with quantum circuits can arise based on the specific location where these operations occur.  In QCCD architectures, these variations are independent of the individual qubits present in the designated zone.  The compiler, responsible for orchestrating quantum circuits, dynamically determines the location for each quantum gate.  Circuits with similar architectures may even have different gate locations as the compiler optimises each circuit to minimise both, transport operations and overall execution time, see e.g.~\cite{Kreppel_Melzer_Millán_Wagner_Hilder_Poschinger_Schmidt-Kaler_Brinkmann}. 

A promising future direction lies in using Rydberg states for trapped ions to realise faster two-qubit entangling gates.  While this idea is still under development, first Rydberg-based trapped-ion qubits have been realised~\cite{higgins2017single}, as well as entangling gates between pairs~\cite{zhang2020submicrosecond}, with substantially faster gate times ($\sim\qty{700}{ns}$) than previous approaches for trapped ions, but a relatively low fidelity of~78\%.  The Rydberg state of an ion is created by exciting the outermost electron into a highly excited state. The resulting large dipole moment enables interactions with other Rydberg-excited ions, which can be used for fast entanglement operations as known from neutral-atom qubits, see~\cite{Wintersperger2023}.  However, due to the large dipole moment, Rydberg states are very sensitive to electric fields and this might introduce Stark shifts or field ionisation, because electric fields are always present due to the trapping potential.

%Further scaling approaches are \cite{Knollmann_etal._2024}

\subsection{Comparison with other platforms\status{KW,OK}}

%% other natural qubits: atoms, photons

\paragraph{Trapped ions and neutral atoms}

Some of the challenges for trapped-ion quantum computers described above are also faced by other platforms, such as neutral atoms: They also suffer from relatively slow single-qubit gate operations. The reason for this is that the gates are also implemented by optical means and thus depend on the properties of atomic transitions.  Atomic qubits can also be moved and arranged in different compute zones~\cite{Bluvstein2023}, which offers flexibility and an effective all-to-all connectivity, but increases the compute time further, similarly as for ion traps.  Compared to neutral atom devices, most trapped-ion quantum computers exhibit better gate fidelities, especially for two-qubit gates, as well as longer coherence times.
%\ahnote{If there are references backing the statements, it would be good to add them here.} 
On the other hand, atomic setups, often based on optical microtraps~\cite{schymik_enhanced_2020, ebadi_quantum_2021}, are considerably easier to scale to larger numbers of qubits. 

\paragraph{Trapped ions and superconducting qubits}

Currently available quantum computers with superconducting qubits are more scaled up, featuring up to~$\sim 1000$ qubits~\cite{IBM_2023}.  Up to a certain point, superconducting chips can leverage existing semiconductor fabrication techniques and thus are compatible with techniques for scaling up classical processors, enabling the production of large numbers of qubits on a chip.  However, increasing the number of qubits on a chip also involves technical challenges: more control wires need to be connected to the chip and routed to the qubits, which increases the amount of heat brought into the cryostat~\cite{joshi2022scaling}.  Possible solutions include modularisation and the use of cryogenic CMOS technologies~\cite{Bishnu_2020}.  Compared to an ion, a superconducting qubit with a typical size (including resonators, filters, wiring) of up to~\qty{1}{mm^2} is relatively large (Google's Sycamore chip had 53 qubits on a chip area of about~\qty{100}{mm^2}~\cite{Arute2019}).
% See also \url{https://quantumcomputing.stackexchange.com/questions/4648/what-limits-scaling-down-the-size-of-superconducting-qubits}}

As mentioned above, ion traps only need cooling to reduce noise and thus do not need to be cooled to the mK-range in contrast to superconducting qubits. Moreover, trapped ions feature better gate fidelities and longer coherence times than superconducting circuits.  Besides, superconducting qubits have a fixed layout and are severely restricted in their qubit-qubit connectivity.  Approaches to extend the connectivity exist, but a higher connectivity also leads to an increase in unwanted interactions between qubits, so-called crosstalk~\cite{stassi_scalable_2020, bravyi_future_2022}.  For trapped-ion setups in turn, crosstalk is not a major issue.  An advantage of superconducting qubits is their fast gate execution times, allowing for much shorter algorithm runtimes. \grnote{Even though the following is prominent in the comparison between ion and superconducting the statement is more maybe more general. New paragraph/line break?}  Thus, the comparison of trapped-ion and superconducting quantum computers is (another) example for the trade-off between high fidelity and high speed, which makes more sense for the quantum computer providers, while for users, the total performance like time-to-solution or total-cost-of-solution should be more important. There are metrics like quantum volume of IBM~\cite{Cross2019} and algorithmic qubits of IonQ~\cite{chen_benchmarking_2023}, attempting to assess the total performance in a technology-independent way.  According to such metrics, currently trapped-ion quantum computers typically score better than superconducting alternatives.

\subsection{Conclusion\status{KW,OK}}
\label{sec:disc_conclusion}

In this work we give an overview of trapped-ion quantum computers, including their main physical properties and operating principles, as well as their advantages and disadvantages with respect to different applications.

Trapped-ion quantum computers have demonstrated tremendous potential for advancing the field of quantum computing, e.\,g., for some NISQ applications they already outperform other quantum computing platforms. The strengths of this technology lie in its high-fidelity qubits, long coherence times and effective all-to-all connectivity.  Moreover, the physical transport of ion qubits over longer distances via shuttling allows to define separate zones, i.\,e., for compute operations and storage, which reduces unwanted crosstalk and plays an important role for the implementation of fault-tolerant quantum computers. 
However, there are challenges to address, including complex and resource-intensive operations, slow gate speeds and the scaling to very large qubit numbers. 

Looking ahead, further research and development is needed to overcome these challenges and unlock the full potential of trapped-ion quantum computers.  Innovations in qubit manipulation techniques, further advances towards scaling, such as the QCCD architecture, and the implementation of fault-tolerant quantum error correction codes hold promise for enhancing the performance and efficiency of trapped-ion systems.  Additionally, advances in system integration, ion trap miniaturisation, and reduction of environmental noise will be important for practical implementation.  Characterising the performance of quantum computers is a critical challenge as these systems increase in size and complexity.

\section*{Acknowledgement\status{TE,OK}}
\label{sec:acknowledgement.tex}

We would like to thank the following trapped-ion quantum computing providers, listed in alphabetical order, for the time they spent with us to discuss their quantum computing platform: AQT, eleQtron, IonQ, neQxt, Oxford Ionics, Quantinuum.

%%%%%%%%%%%%%%%%%%%%%%%%%%%%%%%%%%%%%%%%%%%%%%%%%%%%%%%%%%%%%%%%%%%%%%%%%%%%

\begin{backmatter}

\section*{Competing interests}

The authors declare that they have no competing interests.  Dr. Sebastian Luber is an employee of Infineon Technologies AG, which offers components for trapped-ion quantum computers.

%\section*{Author's contributions}
%    Text for this section \ldots

\section*{List of abbreviations}

\begin{tabular}{c|c}
  \hline
  AOD & acousto-optic deflector \\ \hline
  AOM & acousto-optic modulator \\ \hline
  NISQ  & Noisy intermediate-scale quantum \\ \hline
  QAOA & Quantum Approximate Optimisation Algorithm \\ \hline
  QCCD & Quantum charge-coupled device \\ \hline
  RF & radio frequency\\ \hline
  SPAM & state preparation and measurement \\ \hline
\end{tabular}
  
%%%%%%%%%%%%%%%%%%%%%%%%%%%%%%%%%%%%%%%%%%%%%%%%%%%%%%%%%%%%%
%%                  The Bibliography                       %%
%%                                                         %%
%%  Bmc_mathpys.bst  will be used to                       %%
%%  create a .BBL file for submission.                     %%
%%  After submission of the .TEX file,                     %%
%%  you will be prompted to submit your .BBL file.         %%
%%                                                         %%
%%                                                         %%
%%  Note that the displayed Bibliography will not          %%
%%  necessarily be rendered by Latex exactly as specified  %%
%%  in the online Instructions for Authors.                %%
%%                                                         %%
%%%%%%%%%%%%%%%%%%%%%%%%%%%%%%%%%%%%%%%%%%%%%%%%%%%%%%%%%%%%%

\bibliographystyle{bmc-mathphys} 
\bibliography{literature}

%%%%%%%%%%%%%%%%%%%%%%%%%%%%%%%%%%%
%%                               %%
%% Figures                       %%
%%                               %%
%% NB: this is for captions and  %%
%% Titles. All graphics must be  %%
%% submitted separately and NOT  %%
%% included in the Tex document  %%
%%                               %%
%%%%%%%%%%%%%%%%%%%%%%%%%%%%%%%%%%%

%%
%% Do not use \listoffigures as most will included as separate files

%\section*{Figures}

%%%%%%%%%%%%%%%%%%%%%%%%%%%%%%%%%%%
%%                               %%
%% Abbreviations                 %%
%%                               %%
%%%%%%%%%%%%%%%%%%%%%%%%%%%%%%%%%%%

%% Use of \listoftables is discouraged.

%%%%%%%%%%%%%%%%%%%%%%%%%%%%%%%%%%%
%%                               %%
%% Additional Files              %%
%%                               %%
%%%%%%%%%%%%%%%%%%%%%%%%%%%%%%%%%%%

%\section*{Additional Files}
%  \subsection*{Additional file 1 --- Sample additional file title}
%    Additional file descriptions text (including details of how to
%    view the file, if it is in a non-standard format or the file extension).  This might
%    refer to a multi-page table or a figure.

%  \subsection*{Additional file 2 --- Sample additional file title}
%    Additional file descriptions text.

\end{backmatter}

\end{document}